\documentclass[useAMS,usenatbib,,referee]{biom}

\usepackage{graphicx}
\usepackage{amssymb}
\usepackage[figuresright]{rotating}
\usepackage{subfigure}
\usepackage{amsmath}
\usepackage{enumitem}
\usepackage{multirow}
\usepackage{enumerate}
\usepackage{color}
\usepackage{url}
\usepackage{tabu}
\usepackage{caption}
\usepackage{natbib}    
\usepackage{threeparttable}
\usepackage{chngcntr}
\usepackage{setspace}
\usepackage{mdwlist}
\usepackage{diagbox}
\usepackage{rotating}
\usepackage{float}
\usepackage{setspace}
\usepackage{comment}
\usepackage{titlesec}
\newcommand{\tabincell}[2]{\begin{tabular}{@{}#1@{}}#2\end{tabular}}

\titlespacing*{\section}{0pt}{0.5\baselineskip}{0.2\baselineskip}
\titlespacing*{\subsection}{0pt}{0.2\baselineskip}{0.2\baselineskip}

\allowdisplaybreaks
\title[]{Association study between gene expression and multiple phenotypes in omics applications of complex diseases}

\author{Yujia Li, Yusi Fang, Peng Liu, and George C. Tseng$^{*}$\email{ctseng@pitt.edu} \\
Department of Biostatistics, University of Pittsburgh, Pittsburgh, PA 15261.\\
}

\begin{document}



\pagerange{\pageref{firstpage}--\pageref{lastpage}} 
\volume{64}
\pubyear{2008}
\artmonth{December}


\doi{10.1111/j.1541-0420.2005.00454.x}
\label{firstpage}

\begin{abstract}
Studying phenotype-gene association can uncover mechanism of diseases and develop efficient treatments. In complex disease where multiple phenotypes are available and correlated, analyzing and interpreting associated genes for each phenotype respectively may decrease statistical power and lose intepretation due to not considering the correlation between phenotypes. The typical approaches are many global testing methods, such as multivariate analysis of variance (MANOVA), which tests the overall association between phenotypes and each gene, without considersing the heterogeneity among phenotypes. In this paper, we extend and evaluate two p-value combination methods, adaptive weighted Fisher's method (AFp) and adaptive Fisher's method (AFz), to tackle this problem, where AFp stands out as our final proposed method, based on extensive simulations and a real application. Our proposed AFp method has three advantages over traditional global testing methods. Firstly, it can consider the heterogeneity of phenotypes and determines which specific phenotypes a gene is associated with, using phenotype specific 0-1 weights. Secondly, AFp takes the p-values from the test of association of each phenotype as input, thus can accommodate different types of phenotypes (continuous, binary and count). Thirdly, we also apply bootstrapping to construct a variability index for the weight estimator of AFp and generate a co-membership matrix to categorize (cluster) genes based on their association-patterns for intuitive biological investigations. Through extensive simulations, AFp shows superior performance over global testing methods in terms of type I error control and statistical power, as well as higher accuracy of 0-1 weights estimation over AFz. A real omics application with transcriptomic and clinical data of complex lung diseases demonstrates insightful biological findings of AFp.

\end{abstract}


\maketitle
\section{Introduction}
\label{sec:AW_intro}

Identifying genes associated with phenotypes of interest in transcriptomics studies can help understand mechanism of diseases and extensive efforts have been tried . For example, \cite{peters2015transcriptional} identified 1197 genes associcated with chronological age using whole-blood gene expression data and \cite{blalock2004incipient} tested gene expression with MiniMental Status Examination (MMSE) and neurofibrillary tangle (NFT) in Alzheimer's disease (AD) patients, identifying thousands of genes significantly correlated with two AD markers respectively. One special case is the differentially expressed (DE) analysis where the phenotype of interest is binary (e.g., diease/control). Identifying DE genes and the enriched pathways has become a standard pipeline in modern transcriptomics application and many softwares have been developed \citep{robinson2010edger,ritchie2015limma}. However, a complex disease is usually characterized by multiple phenotypes, reflecting different facets of the disease. Take chronic obstructive pulmonary disease (COPD) as an example, the disease can be characterized by multiple phenotypes, such as FEV1 (i.e., the volume of breath exhaled with effort in one second), FVC (i.e., the full amount of air that can be exhaled with effort in a complete breath), and differential test of white blood cells. A naive way is to test the association between each gene and each single phenotype at a time and interpret each set of associated genes. However, it may decrease the statistical power when each phenotype is weakly associated with a gene and lose intepretation by failing to find a common gene set associated with multiple phenotypes. Therefore, how to jointly analyze the association between a gene and multiple phenotypes is the interest of this paper. Below we introduced four remaining challenges in omics applications, followed by a review of existing methods. 

The first issue is to identify the genes associated with the phenotypes (i.e, global test). We consider the following union-intersection test (UIT) \citep{roy1953heuristic} or conjunction null hypothesis \citep{benjamini2008screening} in the statistical literatures for each gene:
$$
\begin{array}{l}
	H_{0}: \vec{\theta} \in \bigcap\{\theta_k=0\} \\
	H_{A}: \vec{\theta} \in \bigcup\{\theta_k\neq 0\}
\end{array}
$$
to determine the overall association between a gene and $K$ phenotypes of interest, where $\theta_k$ is the effect size for phenotype $k$,  $1 \leq k \leq K$. The existing methods to solve this question can be summarized into three categories. The first category includes some classical methods for multivariate data, such as multivariate analysis of variance (MANOVA), linear mixed models (LMMs) and generalized linear models (GLMM). MANOVA and LMMs require Gaussian assumption and cannot be applicable to non-Gaussian phenotypes (e.g., binary and count-based data). GLMM cannot be used when the phenotypes are mixed with more than one data types. We will evaluate MANOVA in this paper. The second category is regression-based methods. \cite{o2012multiphen} proposed MultiPhen method by regressing genotypes on phenotypes via proportional odds logistic model, which is only applicable to ordinal outcome in GWAS studies. \cite{wu2016sequence} proposed multi-trait gene sequence kernel association test (MSKAT), where the purpose is to test the association between a phenotype and multiple SNPs in a chromesome region and cannot be easily extended to our scenario. Therefore, we don't include these two regression-based methods into evaluation in this paper.

The third category is to combine summary statistics (e.g., p-values and test statistics) from association test of each phenotype, which has been widely used in meta-analysis field to identify DE genes for multiple studies and GWAS setting with multiple phenotypes. These methods can lead to increased statistical power by combining the summary statistics representing the strength of association between a gene and each phenotype. Moreover, methods by combining p-values can accommodate phenotypes with mixed data types since each p-value can be calculated respectively using different approaches. \cite{o1984procedures} combined the test statistics from the individual test on each trait weighted by inverse variance. \cite{pan2014powerful} and \cite{zhang2014testing} proposed the sum of powered score tests (SPU) and adaptive SPU (aSPU) to combine the score test statistics derived from generalized estimation equations (GEE) in GWAS settings. We include aSPU into evaluation with two variants, the working correlation matrix of GEE being diagonal (aSPU.ind) or exchangeable (aSPU.ex), where aSPU.ind was used as default in \cite{zhang2014testing}. Many p-value combination methods were also developed, including Fisher's method  \citep{fisher1992statistical}, the minimum p-value method (minP) \citep{tippett1931methods}, and many others.  Fisher's method defines the test statistic for each gene as the sum of log-transformed p-values by giving all p-values equal weights: $T^{Fisher}=-2 \sum_{k=1}^{K} \log p_k$, where $p_k$ is the p-value from the test of $k$th phenotype. A larger Fisher's score indicates stronger association. minP uses the minimum p-value among $p_1, p_2, \cdots, p_K$ as the test statistic. Under meta-analysis setting, p-values from different studies are independent and the null distribution of Fisher's and minP method are $\chi^{2}_{2K}$ and beta distribution respectively. In our setting, the phenotypes are correlated and the null distribution can be generated by randomly permuting the sample order for gene expression data, keeping the phenotypes data unchanged, which breaks the association between genes and phenotypes while keeping the correlation among phenotypes. \cite{van2013tates} proposed to exploit the correlation among p-values from each univariate test and generate a new test statistic (TATES). We include Fisher, minP, and TATES into evaluation.

The second challenge is to characterize heterogeneity across phenotypes. Take Fisher as an example, suppose $K=3$ , $\vec{p}_{1}=(0.001, 1, 1)$ represents p-values of three phenotypes of Gene 1 and $\vec{p}_{2}=(0.1, 0.1, 0.1)$ represents p-values of Gene 2. Both genes produce the same Fisher's statistics ($T^{Fisher}=13.8$), but the biological interpretation of the two genes are obviously different. $\vec{p}_{1}$ indicates strong statistical significance only in the first phenotype, while $\vec{p}_{2}$ shows marginal statistical significance in all three phenotypes. All the methods mentioned above fail to characterize heterogeneity across phenotypes, and to tackle this problem, adaptive Fisher's method \citep{song2016screening} and adaptive weighted Fisher's method \citep{huo2020p} are two applicable methods. Both methods generate 0-1 weights vector $\vec{w}=(w_1, w_2, w_3)$ where $w_k=1$ ($1 \leq k \leq K$, suppose $K=3$) indicates this phenotype is associated with a gene and 0 otherwise. Both methods are proposed under meta-analysis with independent multiple studies where searching for the optimal weight from the smallest to largest p-value is sufficient. In other words, they ordered p-values ($p_1, p_2, p_3$) from the smallest to largest to get ordered p-values $(p_{(1)}, p_{(2)}, p_{(3)})$ and determined the best $\vec{w}$ from $(1, 0, 0), (1, 1, 0)$ and $(1, 1, 1)$. However, in our setting, phenotypes are correlated and we propose to extend these two methods by considering correlation among phenotypes using combinatorial search with the original $(p_1, p_2, p_3)$ and determine the best $\vec{w}$ among $(1, 0, 0), (0, 1, 0), (0, 0, 1), (1, 0, 1), (0, 1, 1), (1, 1, 0)$ and $(1, 1, 1)$. To simplify notation hereafter, we abbreviate adaptive Fisher's method \cite{song2016screening} as AFz since it utilizes standardized sum statistics as test statistic to determine the optimal weight, and abbreviate adaptive weighted Fisher's method \cite{huo2020p} as AFp since it uses the p-value of sum statistics as the proposed test statistic. The details of AFz and AFp will be introduced in Section \ref{sec:AFp} and \ref{sec:AFz}.

The third issue remained is to estimate the uncertainty of 0-1 weights. \cite{huo2020p} proposed a bootstrapping method to estimate the variability index under independent settings and we borrow the idea in our setting for AFp and AFz. The fourth challenge is to identify clusters of genes as gene modules ($M_1, M_2, \cdots M_q$). For each two genes $i$ and $j$ within the same gene module, $v_i^k=v_j^k$ ($\forall$ $1 \leq k \leq K$), where $v_i^k=w_i^k$sign($\theta_{i}^k$) $ \in \{-1, 0, 1\}$ and sign($\theta_{i}^k$) indicates the up-regulation or down-regulation. Given $K$ studies, the resulting genes could be categorized into ($3^K-1$) groups. This becomes intractable for further biological investigation when $K$ is large. For example, combining $K=5$ studies produces $35-1=242$ categories of biomarkers. Following \cite{huo2020p}, we propose to estimate comembership matrix of genes through bootstrapping, followed by tight clustering method \citep{tseng2005tight}, to tackle this issue.

AFp and AFz can both be used to determine genes associated with phenotypes, characterize heterogeneity among phenotypes, estimate uncertainy of estimated weights and clustering genes into gene modules, while the performance of AFp and AFz hasn't been fully studied. In this paper, we systematically evaluate AFp and AFz. Our contributions are two-fold: 1) We use extensive simulation to evaluate the type I error rate and power for AFp, AFz, minP, Fisher, TATES, aSPU.ind and aSPU.ex where AFp and AFz show robust performance in all scenarios. 2) We comprehensively evaluate the performance of AFp and AFz for characterizing heterogeneity (i.e., accuracy of 0-1 weights estimation) and identify AFp as the best performer. Therefore, AFp is the final method we recommend for general purpose of gene-phenotype association analysis and gene categorization. The article is structured as follows. In Section 2, we introduce AFp and AFz methods respectively under settings of correlated phenotypes (Section \ref{sec:AFp} and \ref{sec:AFz}), followed by the bootstrapping algorithms to estimate variability index and categorize genes (Section \ref{sec:varibility} and \ref{sec:categorization}). Section \ref{sec:simulation} includes extensively simulations to evaluate AFp, AFz, as well as other existing methods. Section \ref{sec:AW_application} contains the result from a lung disease transcriptomic dataset. We include final conclusion and discussion in Section \ref{sec:discussion}.

\section{Methods}
\label{chapter3:methods}
In this section, we will introduce AFp and AFz methods, which can determine whether a gene is associated with all the phenotypes, characterize the heterogeoity among phenotypes by 0-1 weights, estimate variability index of weights estimated and cluster genes based on their association pattern with phenotypes. Suppose there are $n$ independent, $K$ phenotypes, $p$ gene features and $M$ covariates. Denote by $y_{ik}$, $z_{im}$ snd $x_{ij}$ the $k$th phenotype, $m$th covariate and $j$th gene feature of subject $i$ respectively, where $1 \leq k \leq K$, $1 \leq m \leq M$, and $1 \leq j \leq p$. $\overrightarrow{\boldsymbol{Y_k}}=(y_{1k}, y_{2k},\ldots, y_{nk})^{T}$, $\overrightarrow{\boldsymbol{Z_m}}=(z_{1m}, z_{2m}, \ldots, z_{nm})^{T}$, and $\overrightarrow{\boldsymbol{X_j}}=(x_{1j}, x_{2j}, \ldots, x_{nj})^{T}$ are the vectors $k$th phenotype, $m$th covariate or $j$th gene for all the samples. 

\subsection{Generalized linear models}

For both AFp and AFz, a generalized linear model of the following form is assumed for the $k$th phenotype and the $j$th gene with $M$ covariates:
\begin{equation}
	\label{eq:glm}
	g_{k}(E(Y_{i k}))=x_{ij} \cdot \theta_{jk}+\sum_{m=1}^{M} z_{i m} \cdot \alpha_{m k j}
\end{equation}
where $\theta_{jk}$ is the coefficient of gene featture, $\alpha_{m k j}$ is the coefficient of covariates and $g_k()$ is the link function. The associated p-value $p_{jk}$ of $\theta_{jk}$ can be derived by classic Wald test or Score test, which serves as input for AFp and AFz. Considering heterogeity among phenotypes, weighted statistic 
\begin{equation*}
	U_{j}(\overrightarrow{\boldsymbol{w_{j}}})=-\sum_{k=1}^{K} w_{j k} \log (p_{j k})
\end{equation*}
can be used to determine whether $j$th gene is associated with phenotypes, where $w_{j k}\in \{0,1\}$ can tell which phenotypes contribute to the association, $\overrightarrow{\boldsymbol{w_{j}}}=(w_{j1},w_{j2},\ldots, w_{jK})^T$ and the searching space $\Omega=\left\{\overrightarrow{\boldsymbol{w_j}}: \overrightarrow{\boldsymbol{w_j}} \neq 0, \overrightarrow{\boldsymbol{w_j}}=\left(w_{j1}, \ldots, w_{jK}\right) \in\{0,1\}^{K}\right\}$ contains $2^{K}-1$ non-zero vectors of weights (See Section \ref{sec:AFp} and \ref{sec:AFz} for details). The observed weighted statistic of $U_{j}(\overrightarrow{\boldsymbol{w_{j}}})$ is denoted 	as $u_{j}(\overrightarrow{\boldsymbol{w_{j}}})$.

\subsection{Permutation}
\label{sec:permutation}
Both AFp and AFz are derived from observed weighted statistic $u_{j}(\overrightarrow{\boldsymbol{w_{j}}})$, $1\leq j\leq p$. AFp uitilizes the smallest p-value of $u_{j}(\overrightarrow{\boldsymbol{w_{j}}})$ among all possible weight $\overrightarrow{\boldsymbol{w_{j}}}$ as test statistic while AFz standardizes $u_{j}(\overrightarrow{\boldsymbol{w_{j}}})$ based on mean and standard deviation under null (See Section \ref{sec:AFp} and \ref{sec:AFz}).  Since the mean, standard deviation and p-value of observed weighted statistics $u_{j}(w_{j})$ are intractable to calculated analytically, permutation methods will be used to generate null distribution of $U_{j}(\overrightarrow{\boldsymbol{w_{j}}})$. We intend to test the conditional independence between each gene and $K$ phenotypes given the covariates $\boldsymbol{Z}$ (i.e., $\boldsymbol{Y} \perp \overrightarrow{\boldsymbol{X_j}} \mid \boldsymbol{Z}$), where $\boldsymbol{Z}=(\overrightarrow{\boldsymbol{Z_1}}, \overrightarrow{\boldsymbol{Z_2}}, \ldots \overrightarrow{\boldsymbol{Z_m}})$ and $\boldsymbol{Y}=(\overrightarrow{\boldsymbol{Y_1}}, \overrightarrow{\boldsymbol{Y_2}}, \ldots \overrightarrow{\boldsymbol{Y_K}})$. The permutation procedure should break the associations between $\boldsymbol{Y}$ and $\overrightarrow{\boldsymbol{X_j}}$ while preserving the associations between $\boldsymbol{Y}$ and $\boldsymbol{Z}$ and between $\overrightarrow{\boldsymbol{X_j}}$ and $\boldsymbol{Z}$. Simply permuting the genotype $\overrightarrow{\boldsymbol{X_j}}$ leads to inflated type I error rate, because the correlation between $\overrightarrow{\boldsymbol{X_j}}$ and covariates $\boldsymbol{Z}$ are also destroyed. Following \cite{potter2005permutation} and \cite{werft2010glmperm}, we permute residuals of regressions of $\overrightarrow{\boldsymbol{X_j}}$ on $\boldsymbol{Z}$ for generalized regression models. That is,  we first regress each gene feature on the covariates, then permute the residuals derived from the regression and fit the generalized linear model by regressing each phenotype on the permuted residuals.

Specifically, we denote the vector of residuals of regressing $\boldsymbol{X_j}$ on $\boldsymbol{Z}$ as $\overrightarrow{\boldsymbol{e_j}}=(e_{1j}, e_{2j}, \ldots, e_{nj})^T$ and permute it for $B$ times. In the $b$th permutation, we regress $\overrightarrow{\boldsymbol{Y}_{k}}$ on $\boldsymbol{e}_{\boldsymbol{j}}^{(b)}$ by generalized linear model $g_{k}(E(Y_{i k}))=e^{(b)}_{ij} \cdot \theta^{(b)}_{jk}$ ($1 \leq i \leq n$) and get the p-values $p_{jk}^{(b)}$ for the coefficient $\theta^{(b)}_{jk}$. After $B$ permutations, we get a $B \times K$ matrix $\mathbb{P}=\{p_{jk}^{(b)}\}$. The oberved weighted statistics for permuted data can be derived as $u_{j}^{(b)}(\overrightarrow{\boldsymbol{w_{j}}})=-\sum_{k=1}^{K} w_{j k} \log (p^{(b)}_{j k})$. Note that we break the association of each gene and each phenotype. Therefore, for a given weight $\overrightarrow{\boldsymbol{w_{j}}}$, the null distribution of observed weighted statistic $u_{j}(\overrightarrow{\boldsymbol{w_{j}}})$ is $\{p^{(b)}_{j^{\prime} k}, 1\leq b\leq B, 1\leq j^{\prime}\leq p\}$ with precision $B\times p$.

\subsection{AFp}
\label{sec:AFp}
Under the null hypothesis that $\theta_{j k}=0$, $\forall k$. the p-value of observed weighted statistic, $p_{U}(u_{j}(\overrightarrow{\boldsymbol{w_{j}}}))$, can be obtained for $j$th gene and a given weight $\overrightarrow{\boldsymbol{w_{j}}} \in \Omega$ by 
\begin{equation*}
	p_{U}(u_{j}(\overrightarrow{\boldsymbol{w_{j}}}))=\frac{\sum_{b=1}^{B} \sum_{j^{\prime}=1}^{p} I\{u_{j^{\prime}}^{(b)}(\overrightarrow{\boldsymbol{w_{j}}}) \geq u_{j}(\overrightarrow{\boldsymbol{w_{j}}})\}}{B \cdot p}
\end{equation*}
and AFp statistic is defined as the minimal p-value among all possible weights $\overrightarrow{\boldsymbol{w_{j}}} \in \Omega$:
\begin{equation*}
	T_{j}^{\mathrm{AFp}}=\min _{\overrightarrow{\boldsymbol{w_{j}}} \in \Omega} p_{U}(u_{j}(\overrightarrow{\boldsymbol{w_{j}}})).
\end{equation*}
The associate weight vector 
\begin{equation*}
	\overrightarrow{\boldsymbol{w_{j}}}^{AFp}=\arg \min _{\overrightarrow{\boldsymbol{w_{j}}} \in \Omega} p_{U}(u_{j}(\overrightarrow{\boldsymbol{w_{j}}}))
\end{equation*}
can determine heterogeneity of genes (second issue discussed in Section \ref{sec:AW_intro}) and serve as a convenient basis for gene categorization in follow-up biological interpretations and explorations (third and fourth issue discussed in Section \ref{sec:AW_intro}). To get the p-value of $T_{j}^{\mathrm{AFp}}$, we similarly calculate $T_{j}^{\mathrm{AFp, (b)}}$ using $\mathbb{P}=\{p_{jk}^{(b)}\}$ from permutation. Specifically, we calculate
\begin{equation*}
	T_{j}^{\mathrm{AFp, (b)}}=\min _{\overrightarrow{\boldsymbol{w_{j}}} \in \Omega} p_{U}(u^{(b)}_{j}(\overrightarrow{\boldsymbol{w_{j}}})),
\end{equation*}
where $p_{U}(u^{(b)}_{j}(\overrightarrow{\boldsymbol{w_{j}}}))=\frac{\sum_{b=1}^{B} \sum_{j^{\prime}=1}^{p} I\{u_{j^{\prime}}^{(b)}(\overrightarrow{\boldsymbol{w_{j}}}) \geq u^{(b)}_{j}(\overrightarrow{\boldsymbol{w_{j}}})\}}{B \cdot p}$ and the p-value of $T_{j}^{\mathrm{AFp}}$ can be calculated as
\begin{equation*}
	p_{T}(T^{AFp}_{j})=\frac{\sum_{b=1}^{B} \sum_{j^{\prime}=1}^{p} I\{T_{j^{\prime}}^{AFp, (b)} \leq T_{j}^{AFp}\}}{B \cdot p}.
\end{equation*}
In summary, $p_{T}(T^{AFp}_{j})$ can be used to determine whether $j$th gene is associated with $K$ phenotypes and $\overrightarrow{\boldsymbol{w_{j}}}^{AFp}$ can be used to determine which specific phenotypes the $j$th gene is associated with.

\subsection{AFz}
\label{sec:AFz}
For a given $\overrightarrow{\boldsymbol{w_{j}}} \in \Omega$, the mean and standard deviation of $u_{j}(\overrightarrow{\boldsymbol{w_{j}}})$ under null are $E(u_{j}(\overrightarrow{\boldsymbol{w_{j}}}))=\frac{\sum_{b=1}^{B} \sum_{j^{\prime}=1}^{p} u^{(b)}_{j^{\prime}}(\overrightarrow{\boldsymbol{w_{j}}})}{B \cdot p}$
and $sd(u_{j}(\overrightarrow{\boldsymbol{w_{j}}}))=\sqrt{\frac{\sum_{b=1}^{B} \sum_{j^{\prime}=1}^{p} \{u_{j^{\prime}}^{(b)}(\overrightarrow{\boldsymbol{w_{j}}})-E(u_{j}(\overrightarrow{\boldsymbol{w_{j}}}))\}^2}{B \cdot p}}$ . AFz calculates the standarized observed weighted statistic, 
\begin{equation*}
	u^{\prime}_{j}(\overrightarrow{\boldsymbol{w_{j}}})=\frac{u_{j}(\overrightarrow{\boldsymbol{w_{j}}})-E(u_{j}(\overrightarrow{\boldsymbol{w_{j}}}))}{sd(u_{j}(\overrightarrow{\boldsymbol{w_{j}}}))},
\end{equation*}
and the AFz statistic is defined as the largest standarized observed weighted statistic among all possible weights:
\begin{equation*}
	T_{j}^{\mathrm{AFz}}=\max _{\overrightarrow{\boldsymbol{w_{j}}} \in \Omega} u^{\prime}_{j}(\overrightarrow{\boldsymbol{w_{j}}})
\end{equation*}
The associate weight vector can be determined as
\begin{equation*}
	\overrightarrow{\boldsymbol{w_{j}}}^{AFz}=\arg \max _{\overrightarrow{\boldsymbol{w_{j}}} \in \Omega} u^{\prime}_{j}(\overrightarrow{\boldsymbol{w_{j}}})
\end{equation*} 
To calculate p-value of $T_{j}^{\mathrm{AFz}}$, we obtain standardized observed weighted statistic of permuted data by $u^{\prime, (b)}_{j}(\overrightarrow{\boldsymbol{w_{j}}})=\frac{u^{(b)}_{j}(\overrightarrow{\boldsymbol{w_{j}}})-E(u^{(b)}_{j}(\overrightarrow{\boldsymbol{w_{j}}}))}{sd(u^{(b)}_{j}(\overrightarrow{\boldsymbol{w_{j}}}))}$ and $T_{j}^{\mathrm{AFz, (b)}}=\max _{\overrightarrow{\boldsymbol{w_{j}}} \in \Omega} u^{\prime, (b)}_{j}(\overrightarrow{\boldsymbol{w_{j}}})$, where $E(u^{(b)}_{j}(\overrightarrow{\boldsymbol{w_{j}}}))=E(u_{j}(\overrightarrow{\boldsymbol{w_{j}}}))$ and $sd(u^{(b)}_{j}(\overrightarrow{\boldsymbol{w_{j}}}))=sd(u_{j}(\overrightarrow{\boldsymbol{w_{j}}}))$ by definition. Finally, the p-value of $\overrightarrow{\boldsymbol{w_{j}}}^{AFz}$ is calculated as 
\begin{equation*}
	p_{T}(T^{AFz}_{j})=\frac{\sum_{b=1}^{B} \sum_{j^{\prime}=1}^{p} I\{T_{j^{\prime}}^{AFz, (b)} \geq T_{j}^{AFz}\}}{B \cdot p}
\end{equation*}
In summary, $p_{T}(T^{AFz}_{j})$ can be used to determine whether $j$th gene is associated with $K$ phenotypes and $\overrightarrow{\boldsymbol{w_{j}}}^{AFz}$ can be used to determine which specific phenotypes the $j$th gene is associated with.

\subsection{Variability index of adaptive weights}
\label{sec:varibility}
The weight estimate $\widehat{\boldsymbol{w}}_{j}=\left(\widehat{w}_{j 1}, \ldots, \widehat{w}_{j K}\right)$ is binary and discontinuous as a function of the input p-values and thus may not be stable. Following \cite{huo2020p}, we use a bootstrap procedure to calculate an estimate of variability index $U_{j k}=4 \cdot \operatorname{Var}\left(\widehat{w}_{j k}\right)$ for $j$th gene and $k$th phenotype, where the normalization factor 4 scales $\widehat{w}_{j k}$ to $[0, 1]$. We obtain $L$ bootstrap samples with $\boldsymbol{\overrightarrow{Y_k}^{(l)}}$, $\boldsymbol{\overrightarrow{Z_m}^{(l)}}$ and $\boldsymbol{\overrightarrow{X_j}^{(l)}}$ for $k$th phenotype, $m$th covariate, $j$th gene, where $1 \leq k \leq K$, $1 \leq m \leq M$, $1 \leq j \leq G$, $l$ is bootstraping index and $1 \leq l \leq L$. Following the same procedure in Section \ref{sec:AFp} and \ref{sec:AFz}, weight estimates for AFp and AFz can be estimated as $\overrightarrow{\boldsymbol{w_{j}}}^{AFp,(l)}=(w^{AFp,(l)}_{j1},\ldots, w^{AFp,(l)}_{jK})$ and $\overrightarrow{\boldsymbol{w_{j}}}^{AFz,(l)}=(w^{AFz,(l)}_{j1},\ldots, w^{AFz,(l)}_{j})$ for $l$th bootstrap and $j$th gene. The final variability indice are obtained by 
\begin{equation*}
	\widehat{U}^{AFp}_{j k}=\frac{4}{L} \sum_{l=1}^{L}\left(\widehat{w}_{j k}^{AFp,(l)}-\frac{1}{L} \sum_{l^{\prime}=1}^{L} \widehat{w}_{j k}^{AFp,(l)}\right)^{2}
\end{equation*}
and
\begin{equation*}
	\widehat{U}^{AFz}_{j k}=\frac{4}{L} \sum_{l=1}^{L}\left(\widehat{w}_{j k}^{AFz,(l)}-\frac{1}{L} \sum_{l^{\prime}=1}^{L} \widehat{w}_{j k}^{AFz,(l)}\right)^{2}
\end{equation*}
respectively.

\subsection{Ensemble clustering for biomarker categorization}
\label{sec:categorization}
Following \cite{huo2020p}, to identify clusters of genes as gene modules ($M_1, M_2, \cdots M_q$) for genes with stable weight estimates, we cluster genes by a co-membership matrix for all pairs of genes where each element of the co-membership matrix represents a similarity of signed weight $\hat{v}=\hat{w}\times$ sign($\hat{\theta}$) of two genes. Similar to Section \ref{sec:varibility}, we bootstrap data $L$ times and obtain the signed weight statistics $\widehat{v_{jk}}^{AFp,(l)}=\widehat{w_{jk}}^{AFp,(l)}\times\text{sign}(\widehat{\theta_{jk}}^{AFp,(l)})$ and  $\widehat{v_{jk}}^{AFz,(l)}=\widehat{w_{jk}}^{AFz,(l)}\times\text{sign}(\widehat{\theta_{jk}}^{AFz,(l)})$ for $j$th gene, $k$th phenotype and $l$th bootstrapping data for AFp and AFz respectively. We next calculate the comembership matrix for $l$th bootstrapping data of AFp as $\boldsymbol{V}^{AFp, (l)} \in \mathbb{R}^{p \times p}$, where $\boldsymbol{V}^{AFp,(l)}_{jj^\prime}=1$ if $\widehat{v}_{jk}{ }^{AFp, (l)}=\widehat{v}_{j^{\prime} k}{ }^{AFp, (l)}$ for all $k$, and $\boldsymbol{V}^{AFp, (l)}_{jj^\prime}=0$ otherwise. The final comembership matrix can be calculated as $\boldsymbol{V}^{AFp}=\sum_{l=1}^{L} \boldsymbol{V}^{AFp, (l)} / L$ and any classic clustering algorithm can be applied to obtain gene categorization. In this paper, we apply tightc clustering method \cite{tseng2005tight} in real applications, which can eliminate the distraction of scattered genes and construct compact gene modules. Similarly, $\boldsymbol{V}^{AFz}$ can be obtained, followed by the tight clustering algorithm.

\section{Simulation}
\label{sec:simulation}
In this section, we conduct three simulations to compare the following: 1) Type I error control and power for AFp, AFz and other existing methods. 2) Accuracy of weight estimate between AFp and AFz. The methods evaluated include MANOVA, aSPU.ind, aSPU.ex, TATES, Fisher, minP, AFp and AFz. Simulation I and II are settings of continuous phenotypes without and with confounders respectively and all methods above can be evaluated. In simulation III, the phenotypes are settings with mixture of count and continuous phenotypes and we benchmark the performance of TATES, Fisher, minP, AFp and AFz. We have two different settings, A and B in each of Simulation I, II and III, where A mimics the scenarios where each phenotype-gene association has similar effect size and B generates the scenarios where some phenotypes have much stronger association with genes compared with other phenotypes. In each simulation setting, we adapt a random effect model to simulate heirachical association structure between 10 phenotypes and 150 genes, where phenotypes $1 \sim 4$ are associated with gene $1\sim 50$, phenotypes $5 \sim 9$ are associated with gene $1\sim 100$ and phenotype $10$ is associated with gene $101\sim 150$. The details of each simulation setting is illustrated below:


\subsection{Simulation Settings}
\subsubsection{Simulation IA and IB:}
Simulation I simulates continuous phenotypes without confounders.
\begin{itemize}
	\item Simulate $u_{i1}, u_{i2}, u_{i3}\sim N(0, \sigma^2_{\mu})$ for each sample,
	where $N()$ stands for Gaussian distribution and $1 \leq i \leq N_1$. $N_1$ is the sample size.
	\item Simulate 10 phenotypes, where $y_{ik}\sim N(u_{i1}, \sigma^2_k)$ for  $1 \leq k \leq 4$, $y_{ik} \sim N(u_{i1}+u_{i2}, \sigma^2_k)$ for $5 \leq k \leq 9$ and  $y_{i10} \sim N(u_{i3},\sigma^2_{10})$.
	\item Simulate 150 gene features, where $x_{ij} \sim N(u_{i1}, \sigma_x^2)$ for $1 \leq j \leq 50$, $x_{ij} \sim N(u_{i2}, \sigma_x^2)$ for $51 \leq j \leq 100$ and $x_{ij} \sim N(u_{i3}, \sigma_x^2)$ for $101 \leq j \leq 150$.
\end{itemize}
We set $N_1=100$, $\sigma_x=0.5$, and $\sigma_{\mu}\in \{0, 0.4, 0.6\}$ corresponds to different effect size. When $\sigma_{\mu}=0$, all the phenotypes are independent from gene features and the larger the $\sigma_{\mu}$ is, the larger association between phenotypes and genes features. For $\sigma_k$, we set two different scenarios. In Simulation IA, we set $\sigma_k=2$ for $1 \leq k \leq 9$ and $\sigma_{10}=1$, where each phenotype-gene association has similar effect size. In Simulation IB, we set $\sigma_1=\sigma_5=0.05$, $\sigma_{10}=1$ and $\sigma_{k}=2$ otherwise, where $\sigma_1=\sigma_5=0.05$ ensures that the first phenotype has much significant association with genes $1\sim 50$ compared with phenotypes $2\sim 9$ and the $5$th phenotype has much significant association with genes $51\sim 100$ compared with phenotypes $5\sim 9$. We use Simulation IB to evaluate the performance when some phenotypes have much stronger association with genes compared with other phenotypes.

\subsubsection{Simulation IIA and IIB:}
Simulation II simulates continuous phenotypes with a confounder $z$ for gense $1\sim 50$ and phenotypes $1\sim 9$.
\begin{itemize}
	\item Simulate $u_{i1}, u_{i2}, u_{i3}\sim N(0, \sigma^2_{\mu})$ and $z_{i}\sim N(0, \sigma_c^2)$
	where $N()$ stands for Gaussian distribution and $1 \leq i \leq N_1$. $N_1$ is the sample size.
	\item Simulate 10 phenotypes, where $y_{ik}\sim N(u_{i1}+z_i, \sigma^2_k)$ for  $1 \leq k \leq 4$, $y_{ik} \sim N(u_{i1}+u_{i2}+z_i, \sigma^2_k)$ for $5 \leq k \leq 9$ and  $y_{i10} \sim N(u_{i3},\sigma^2_{10})$.
	\item Simulate gene expression data for 150 genes, where $x_{ij} \sim N(u_{i1}+z_i, \sigma_x^2)$ for $1 \leq j \leq 50$, $x_{ij} \sim N(u_{i2}, \sigma_x^2)$ for $51 \leq j \leq 100$ and $x_{ij} \sim N(u_{i3}, \sigma_x^2)$ for $101 \leq j \leq 150$.
\end{itemize}
Similar as Simulation I, we set $N_1=100$, $\sigma_x=0.5$ and $\sigma_{\mu}\in \{0, 0.4, 0.6\}$. In Simulation IIA, we set $\sigma_k=2$ for $1 \leq k \leq 9$ and $\sigma_{10}=1$ and in Simulation IIB, we set $\sigma_1=\sigma_5=0.05$, $\sigma_{10}=1$ and $\sigma_{k}=2$ otherwise.

\subsubsection{Simulation IIIA and IIIB:}
Simulation III generates phenotypes with mixture of count and continuout data. 
\begin{itemize}
	\item Simulate $u_{i1}, u_{i2}, u_{i3}\sim N(0, \sigma^2_{\mu})$ where $N()$ stands for Gaussian distribution and $1 \leq i \leq N_1$. $N_1$ is the sample size.
	\item Simulate 10 phenotypes, where $y_{ik}\sim Possion(u_{i1})$ for  $1 \leq k \leq 4$, $y_{ik}, \sim N(u_{i1}+u_{i2}, \sigma^2_k)$ for $5 \leq k \leq 9$ and  $y_{i10} \sim N(u_{i3},\sigma^2_{10})$.
	\item Simulate gene expression data for 150 genes, where $x_{ij} \sim N(u_{i1}, \sigma_x^2)$ for $1 \leq j \leq 50$, $x_{ij} \sim N(u_{i2}, \sigma_x^2)$ for $51 \leq j \leq 100$ and $x_{ij} \sim N(u_{i3}, \sigma_x^2)$ for $101 \leq j \leq 150$.
\end{itemize}
Similar as Simulation I and II, we set $N_1=100$, $\sigma_x=0.5$ and $\sigma_{\mu}\in \{0, 0.4, 0.6\}$. In Simulation IIIA, we set $\sigma_k=2$ for $5 \leq k \leq 9$ and $\sigma_{10}=1$ and in Simulation IIIB, we set $\sigma_{5}=0.01$, $\sigma_{10}=1$ and $\sigma_{k}=2$ for $6 \leq k \leq 9$.

\subsection{Benchmark for evaluation}
In Simulation I and II, the phenotypes are continuous and we evaluate MANOVA, aSPU.ind, aSPU.ex, TATES, Fisher, minP, AFp and AFz in terms of Type I error ($\sigma_\mu=0$), power ($\sigma_\mu=0.4, 0.6$) and evaluate AFp and AFz for the accuracy of weight estimation. The type I error and power are calculated by $\frac{\sum_{s=1}^{S}\sum_{j=1}^pI\{p^{(s)}_j<0.05\}}{p\times S}$, where $S=500$ is the number of simulated data for each setting, $p^{(s)}_j$ is the p-value of $j$th gene and $s$th simulated data of a generic method discussed in this paper and $I\{.\}$ is indicator function. For accuracy of weight estimation of AFp and AFz, sensitivity $\frac{\sum_{s=1}^{S}\sum_{j=1}^G\sum_{k=1}^K  \widehat{w}^{(s)}_{jk}I\{w_{jk}=1\}}{\sum_{s=1}^{S}\sum_{j=1}^p\sum_{k=1}^K I\{w_{jk}=1\}}$(The proportion of weights estimated to be 1 when the truth is 1), specificity $\frac{\sum_{s=1}^{S}\sum_{j=1}^p\sum_{k=1}^K  (1-\widehat{w}^{(s)}_{jk})I\{w_{jk}=0\}}{\sum_{s=1}^{S}\sum_{j=1}^p\sum_{k=1}^K I\{w_{jk}=0\}}$ (The proportion of weights estimated to be 0 when the truth is 0) are used for evaluation. As you will see in Section \ref{sec:AW_simulation_result}, AFz method has much worse sensitivity when the effect size for each phenotype is imbalanced (Simulation IB, IIB and IIIB). We also include average weight estimate for each phenotype and genes $1\sim50$, $51\sim100$ and $101\sim 150$ for further inspection (See Section \ref{sec:AW_simulation_result} and Table \ref{table:weight} for details).

In Simulation III, the phenotypes are a mixture of count and continuous data and MANOVA, aSPU.ind and aSPU.ex cannot be used. Therefore, we only evaluate TATES, Fisher, minO, AFp and AFz in Simulation III. The benchmark criterion in Simulation III are the same as that in Simulation I and II.
\subsection{Simulation results}
\label{sec:AW_simulation_result}

Table \ref{table:AW_Sim1_main} shows the type I error, power, sensitivity and specificity for Simulation I. All the methods control the type I error well and AFp and AFz method generally perform among the best in terms of power. For example, in Simulation IA, all the phenotype-gene association has similar effect size and AFp (0.9) and AFz (0.89) have higher power than Fisher (0.86) and MANOVA (0.84) when $\sigma_\mu=0.6$. In Simulation IB, AFp and AFz have power 0.96 when $\sigma_\mu=0.6$, which is similar as minP (0.97) and much higher than Fisher (0.88). In terms of weight estimation, AFp has better sensitivity than AFz and the gap is more significant in Simulation IB (0.48 and 0.77 for AFp compared with 0.18 and 0.20 for AFz). To dig further, in Table \ref{table:weight}, we calculate the average weight estimate of AFp and AFz for each phenotype and 50 genes in Simulation IB under $\sigma_\mu=0.6$. AFz only assigns weight 1 for phenotype 1 for genes $1\sim50$ while AFp also assigns weight 1 to phenotype 2, 3 and 4 (with proportion 0.73). For genes $51\sim100$, AFz almost only has weight 1 in phenotype 5 (proportion of weight 1 on phenotype 6$\sim$9 are 0.04, 0.04, 0.03 and 0.04) while AFp assigns weight 1 to phenotype 6$\sim$9 with probability 0.69, 0.7, 0.69 and 0.69. This means that when a gene has different effect size of association with several phenotypes, AFz will assign weight 1 almost only to the phenotype that has strongest association with the gene, while AFp can assign weight 1 to all associated phenotypes more evenly.

Table \ref{table:AW_Sim2_main} shows the result of Simulation II. MANOVA cannot control type I error well when there are confounders, while all other methods can control type I error well. Similar as Simulation I, AFp and AFz generally perform among the best in terms of power and AFp has better sensitivity than AFz, especially when gene-phenotype association is imbalanced (Table \ref{table:weight}).

Table \ref{table:AW_Sim3_main} summarizes the result in Simulation III when the phenotypes have both count and continuous data. In terms of power, AFp and AFz outperform the other four methods. For example, in Simulation IIIA, the power of TATES, minP, Fisher, AFz and AFp are \{0.16, 0.57, 0.61, 0.64, 0.64\} and \{0.49, 0.94, 0.93, 0.98, 0.98\} for $\sigma_{\mu}=0.4$ and $\sigma_{\mu}=0.6$ respectively. Similar as Simulation I and II, AFp has much better sensitivity in terms of weight estimation than AFz, where AFz almost only assigns weight 1 to the phenotype that has strongest association with the gene (Table \ref{table:weight}).
\begin{table}[!thbp]
	\caption{The result of Simulation IA and IB. For $\sigma_\mu=0$, type I error is shown and since there is no association between genes and phenotypes (i.e., true weights are all 0), sensitivity and specificity of weight estimation is omitted. For $\sigma_\mu=0.6$ and 0.8, power, sensitivity and specificity are shown.}
	\label{table:AW_Sim1_main}
	\begin{tabular}{|l|l|l|l|l|l|l|l|}
		\hline
		\multirow{2}{*}{Benchmark}           & \multirow{2}{*}{Method} & \multicolumn{3}{l|}{Simulation IA} & \multicolumn{3}{l|}{Simulation IB} \\ \cline{3-8} 
		&                         & $\sigma_\mu$=0     & $\sigma_\mu$=0.4    & $\sigma_\mu$=0.6     & $\sigma_\mu$=0     & $\sigma_\mu$=0.4     & $\sigma_\mu$=0.6   \\ \hline
		\multirow{8}{*}{\tabincell{c}{power\\ \&\\type I\\error}}& MANOVA                  & 0.05     & 0.36       & 0.84       & 0.05     & 0.76       & 0.93       \\ \cline{2-8} 
		& aSPU.ind                & 0.05     & 0.42       & 0.67       & 0.05     & 0.46       & 0.69       \\ \cline{2-8} 
		& aSPU.ex                 & 0.04     & 0.41       & 0.67       & 0.05     & 0.44       & 0.69       \\ \cline{2-8} 
		& TATES                   & 0.05     & 0.12       & 0.32       & 0.05     & 0.12       & 0.32       \\ \cline{2-8} 
		& minP                    & 0.05     & 0.32       & 0.83       & 0.05     & 0.77       & 0.97       \\ \cline{2-8} 
		& Fisher                      & 0.05     & 0.43       & 0.86       & 0.05     & 0.71       & 0.88       \\ \cline{2-8} 
		& AFz                     & 0.05     & 0.39       & 0.89       & 0.05     & 0.77       & 0.96       \\ \cline{2-8} 
		& AFp                     & 0.05     & 0.41       & 0.9        & 0.05     & 0.76       & 0.96       \\ \hline
		\multirow{2}{*}{Sensitivity}            & AFz                     & -        & 0.39       & 0.58       & -        & 0.18       & 0.2        \\ \cline{2-8} 
		& AFp                     & -        & 0.45       & 0.73       & -        & 0.48       & 0.77       \\ \hline
		\multirow{2}{*}{Specificity}            & AFz                     & -        & 0.92       & 0.97       & -        & 0.97       & 0.99       \\ \cline{2-8} 
		& AFp                     & -        & 0.86       & 0.9        & -        & 0.9        & 0.91       \\ \hline
	\end{tabular}
\end{table}

\begin{table}[]
	\caption{The result of Simulation IIA and IIB. For $\sigma_\mu=0$, type I error is shown and since there is no association between genes and phenotypes (i.e., true weights are all 0), sensitivity and specificity of weight estimation is omitted. For $\sigma_\mu=0.6$ and 0.8, power, sensitivity and specificity are shown. }
	\label{table:AW_Sim2_main}
	\begin{tabular}{|l|l|l|l|l|l|l|l|}
		\hline
		\multirow{2}{*}{Benchmark}              & \multirow{2}{*}{Method} & \multicolumn{3}{l|}{Simulation IIA} & \multicolumn{3}{l|}{Simulation IIB} \\ \cline{3-8} 
		&                         & $\sigma_\mu$=0     & $\sigma_\mu$=0.4     & $\sigma_\mu$=0.6     & $\sigma_\mu$=0     & $\sigma_\mu$=0.4     & $\sigma_\mu$=0.6   \\ \hline
		\multirow{8}{*}{Power and type I error} & MANOVA                  & 0.35     & 0.53       & 0.85       & 0.78     & 0.78       & 0.94       \\ \cline{2-8} 
		& aSPU.ind                & 0.05     & 0.42       & 0.67       & 0.05     & 0.45       & 0.69       \\ \cline{2-8} 
		& aSPU.ex                 & 0.05     & 0.41       & 0.67       & 0.05     & 0.44       & 0.7        \\ \cline{2-8} 
		& TATES                   & 0.05     & 0.12       & 0.32       & 0.05     & 0.12       & 0.32       \\ \cline{2-8} 
		& minP                    & 0.05     & 0.33       & 0.84       & 0.05     & 0.77       & 0.97       \\ \cline{2-8} 
		& Fisher                      & 0.05     & 0.43       & 0.86       & 0.05     & 0.71       & 0.88       \\ \cline{2-8} 
		& AFz                     & 0.05     & 0.39       & 0.89       & 0.05     & 0.77       & 0.97       \\ \cline{2-8} 
		& AFp                     & 0.05     & 0.42       & 0.9        & 0.05     & 0.77       & 0.96       \\ \hline
		\multirow{2}{*}{Sensitivity}            & AFz                     & -        & 0.39       & 0.57       & -        & 0.18       & 0.2        \\ \cline{2-8} 
		& AFp                     & -        & 0.45       & 0.72       & -        & 0.48       & 0.76       \\ \hline
		\multirow{2}{*}{Specificity}            & AFz                     & -        & 0.92       & 0.97       & -        & 0.97       & 0.99       \\ \cline{2-8} 
		& AFp                     & -        & 0.86       & 0.9        & -        & 0.9        & 0.91       \\ \hline
	\end{tabular}
\end{table}

\begin{table}[!thbp]
	\caption{The result of Simulation IIIA and IIIB. For $\sigma_\mu=0$, type I error is shown and since there is no association between genes and phenotypes (i.e., true weights are all 0), sensitivity and specificity of weight estimation is omitted. For $\sigma_\mu=0.6$ and 0.8, power, sensitivity and specificity are shown. }
	\label{table:AW_Sim3_main}
	\begin{tabular}{|l|l|l|l|l|l|l|l|}
		\hline
		\multirow{2}{*}{Benchmark}              & \multirow{2}{*}{Method} & \multicolumn{3}{l|}{Simulation IIIA} & \multicolumn{3}{l|}{Simulation IIIB} \\ \cline{3-8} 
		&                         & $\sigma_\mu$=0     & $\sigma_\mu$=0.4    & $\sigma_\mu$=0.6     & $\sigma_\mu$=0     & $\sigma_\mu$=0.4     & $\sigma_\mu$=0.6   \\ \hline
		\multirow{5}{*}{\tabincell{c}{power\\ \&\\type I\\error}} & TATEs                   & 0.05     & 0.16       & 0.49       & 0.05     & 0.16       & 0.49       \\ \cline{2-8} 
		& minP                    & 0.05     & 0.57       & 0.94       & 0.05     & 0.81       & 0.99       \\ \cline{2-8} 
		& Fisher                      & 0.05     & 0.61       & 0.93       & 0.05     & 0.75       & 0.94       \\ \cline{2-8} 
		& AFz                     & 0.05     & 0.64       & 0.98       & 0.05     & 0.83       & 1          \\ \cline{2-8} 
		& AFp                     & 0.05     & 0.64       & 0.98       & 0.05     & 0.82       & 0.99       \\ \hline
		\multirow{2}{*}{Sensitivity}            & AFz                     & -        & 0.47       & 0.64       & -        & 0.28       & 0.4        \\ \cline{2-8} 
		& AFp                     & -        & 0.61       & 0.91       & -        & 0.62       & 0.92       \\ \hline
		\multirow{2}{*}{Specificity}            & AFz                     & -        & 0.94       & 0.99       & -        & 0.97       & 1          \\ \cline{2-8} 
		& AFp                     & -        & 0.87       & 0.89       & -        & 0.89       & 0.89       \\ \hline
	\end{tabular}
\end{table}

\begin{table}[]
	\caption{The average weight estimate for AFp and AFz methods for Simulation IB, IIB and IIIB when $\sigma_{\mu}=0.6$. For example, for AFp method, the average weight estimate for gene 1-50 and $2$nd phenotype $\boldsymbol{Y_2}$ is 0.73 among 500 simulated data. Since the weight can be either 0 or 1, it means 73\% of the weights for $\boldsymbol{Y_2}$ and gene 1-50 is estimated to be 1 among 500 simulated datasets.}
	\label{table:weight}
	\resizebox{\textwidth}{!}{
	\begin{tabular}{|l|l|l|l|l|l|l|l|l|l|l|l|l|}
		\hline
	Simulation                   & Method               & Gene sets    & $\boldsymbol{Y_1}$   & $\boldsymbol{Y_2}$   & $\boldsymbol{Y_3}$   & $\boldsymbol{Y_4}$   & $\boldsymbol{Y_5}$   & $\boldsymbol{Y_6}$   & $\boldsymbol{Y_7}$   & $\boldsymbol{Y_8}$   & $\boldsymbol{Y_9}$   & $\boldsymbol{Y_{10}}$  \\ \hline
	
		\multirow{6}{*}{Simulation IB} & \multirow{3}{*}{AFp} & $\boldsymbol{X_{1}- X_{50}}$    & 1    & 0.73 & 0.73 & 0.73 & 1    & 0.69 & 0.7  & 0.69 & 0.69 & 0.1  \\ \cline{3-13} 
		&                      & $\boldsymbol{X_{51}- X_{100}}$  & 0.08 & 0.1  & 0.1  & 0.1  & 1    & 0.66 & 0.66 & 0.65 & 0.64 & 0.09 \\ \cline{3-13} 
		&                      & $\boldsymbol{X_{101}- X_{150}}$ & 0.08 & 0.09 & 0.09 & 0.1  & 0.05 & 0.09 & 0.09 & 0.09 & 0.09 & 0.99 \\ \cline{2-13} 
		& \multirow{3}{*}{AFz} & $\boldsymbol{X_{1}- X_{50}}$    & 1    & 0    & 0    & 0    & 0.01 & 0    & 0    & 0    & 0    & 0    \\ \cline{3-13} 
		&                      & $\boldsymbol{X_{51}- X_{100}}$  & 0    & 0    & 0    & 0    & 1    & 0.02 & 0.02 & 0.02 & 0.02 & 0    \\ \cline{3-13} 
		&                      & $\boldsymbol{X_{101}- X_{150}}$ & 0.02 & 0.02 & 0.02 & 0.02 & 0.02 & 0.02 & 0.02 & 0.02 & 0.02 & 0.99 \\ \hline
		
				\multirow{6}{*}{Simulation IIB} & \multirow{3}{*}{AFp} & $\boldsymbol{X_{1}- X_{50}}$    & 1    & 0.7  & 0.72 & 0.7  & 1    & 0.67 & 0.68 & 0.7  & 0.7  & 0.1  \\ \cline{3-13} 
		&                      & $\boldsymbol{X_{51}- X_{100}}$  & 0.07 & 0.1  & 0.08 & 0.1  & 1    & 0.67 & 0.66 & 0.66 & 0.64 & 0.1  \\ \cline{3-13} 
		&                      & $\boldsymbol{X_{101}- X_{150}}$ & 0.07 & 0.09 & 0.08 & 0.11 & 0.07 & 0.09 & 0.09 & 0.08 & 0.08 & 0.99 \\ \cline{2-13} 
		& \multirow{3}{*}{AFz} & $\boldsymbol{X_{1}- X_{50}}$   & 1    & 0    & 0    & 0    & 0.01 & 0    & 0    & 0    & 0    & 0    \\ \cline{3-13} 
		&                      & $\boldsymbol{X_{51}- X_{100}}$  & 0    & 0    & 0    & 0    & 1    & 0.02 & 0.02 & 0.02 & 0.02 & 0    \\ \cline{3-13} 
		&                      & $\boldsymbol{X_{101}- X_{150}}$ & 0.02 & 0.02 & 0.02 & 0.03 & 0.02 & 0.02 & 0.02 & 0.02 & 0.02 & 0.99 \\ \hline
		
			\multirow{6}{*}{Simulation IIIB} & \multirow{3}{*}{AFp} & $\boldsymbol{X_{1}- X_{50}}$   & 1    & 1    & 1    & 1    & 1    & 0.87 & 0.88 & 0.88 & 0.86 & 0.1  \\ \cline{3-13} 
		&                      & $\boldsymbol{X_{51}- X_{100}}$ & 0.16 & 0.15 & 0.16 & 0.14 & 0.99 & 0.82 & 0.83 & 0.83 & 0.83 & 0.1  \\ \cline{3-13} 
		&                      & $\boldsymbol{X_{101}- X_{150}}$ & 0.13 & 0.13 & 0.11 & 0.12 & 0.06 & 0.1  & 0.09 & 0.09 & 0.08 & 1    \\ \cline{2-13} 
		& \multirow{3}{*}{AFz} & $\boldsymbol{X_{1}- X_{50}}$   & 0.65 & 0.64 & 0.64 & 0.65 & 0.98 & 0.1  & 0.1  & 0.1  & 0.09 & 0    \\ \cline{3-13} 
		&                      & $\boldsymbol{X_{51}- X_{100}}$  & 0    & 0    & 0    & 0    & 1    & 0.03 & 0.02 & 0.03 & 0.03 & 0    \\ \cline{3-13} 
		&                      & $\boldsymbol{X_{101}- X_{150}}$ & 0    & 0    & 0.01 & 0    & 0.01 & 0    & 0    & 0    & 0.01 & 1    \\ \hline
	\end{tabular}
}
\end{table}

\section{Real application}
\label{sec:AW_application}
We apply MANOVA, aSPU.ind, aSPU.ex, TATES, Fisher, minP, AFp and AFz to a lung disease transcriptomic dataset with originally 319 patients, where majority of patients were diagnosed by two most representative lung disease subtypes: chronic obstructive pulmonary disease (COPD) and interstitial lung disease (ILD). Gene expression data are collected from Gene Expression Omnibus (GEO) GSE47460 and clinical information obtained from Lung Genomics Research Consortium (\url{https://ltrcpublic.com/}). In this paper, fev1\%prd, fvc\%prd, ratiopre, WBCDIFF1 and WBCDIFF4 are five phenotypes of interest. Fev1 (Forced expiratory volume in 1 second) is the volume of air that can forcibly be blown out in first 1 second after full inspiration, and fev1\%prd indicates a person's measured FEV1 normalized by the predicted FEV1 with healthy lung. FVC (Forced vital capacity) is the volume of air that can forcibly be blown out after full inspiration, and fvc\%prd measures FVC normalized by the predicted FVC with healthy lung. Ratiopre is the ratio of FEV1 to FVC and WBCDIFF1 and WBCDIFF4 are blood tests WBC differential neutrophilic(\%) and blood tests WBC differential eosinophils(\%) respectively. Age, gender and BMI are included as confounding covariates $X$ in the Equation \eqref{eq:glm} to calculate the input p-values for Fisher, minP, AFp and AFz. After filtering samples with missing covariates, the final preprocessed dataset contains $N=279$ samples and $p=15,966$ genes. We first evaluate MANOVA, aSPU.ind, aSPU.ex, TATES, Fisher, minP, AFp and AFz based on number of significant genes and then focus on exploring gene categorization for AFp and AFz.

\begin{figure}[!b]
	
	\includegraphics[width=16cm,height=12cm]{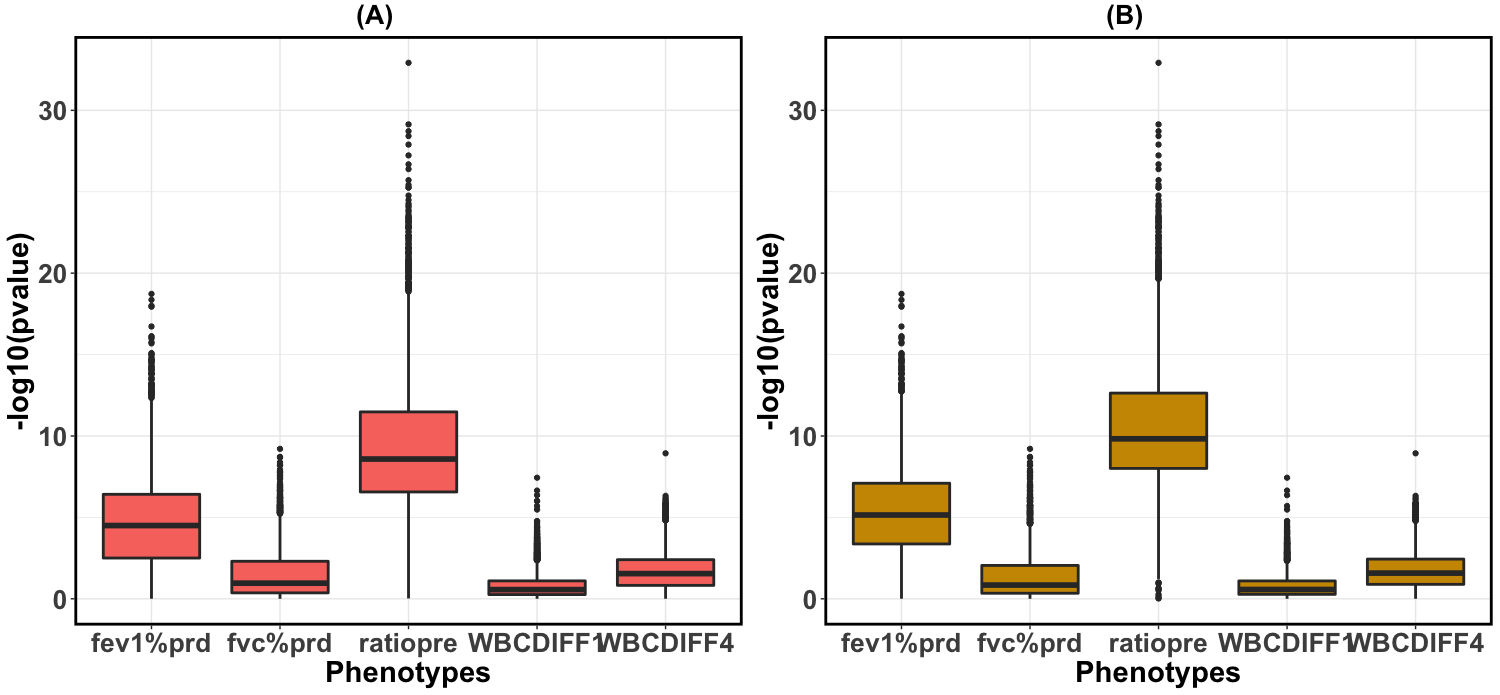}
	\caption{\hspace{10pt}Boxplot of -log10(P-value) of significant genes for each phenotype. (A) shows the significant genes identified by AFp and (B) shows the significant genes determined by AFz.}
	\label{fig:boxplot}
\end{figure}

After calculating the p-value of each gene for each method above, the significant genes are determined by bonferroni correction with cutoff 0.05. We find that aSPU.ex has the most number of signficant genes (6092), followed by AFp (4367), MANOVA (3973), EW (3480), TATES (3443), AFz (3287), Minp (3220) and aSPU.ind (2292). Next. we focus on the significant genes of AFp and AFz methods and try to categorize genes into gene modules. Table \ref{table:weight} shows the precentage of weights estimated to be 1 for 4367 and 3287 significant genes of AFp and AFz respectively, where AFp has a more balanced distribution of weight estimation in all phenotypes and AFz almost only gives weight 1 to ratiopre and gives weight 0 to all the other phenotypes. For example, the percentage of weight 1 for fev1\%prd, fvc\%prd and WBCDIFF1 is 79\%, 48\% and 29\% for AFp while 14\%, 8\% and 1\% for AFz. Figure \ref{fig:boxplot} shows the boxplot of -log10(p-value) of each phenotype for significant genes determined by AFp and AFz methods respectively and it clearly indicates that the p-value of ratiopre is on average much smaller than that of other phenotypes. AFz method almost only gives weight 1 to ratiopre, which is consistent with the findings in Simulation IB, IIB and IIIB (Table \ref{table:weight}). Due to the reason that AFz doesn't assign enough weight 1 to phenotypes except ratiopre,  the result of AFz cannot be further used to do gene categorization, and therefore, we only focus on gene categorization of AFp method from hereafter.

\begin{figure}[!thbp]
	\caption{The heatmap of comembership matrix of seven clusters identifed. Red color means two genes are close. The number in the parentheses indicates the sample size of each cluster. }
	\vspace{2em}
	\includegraphics[width=16cm,height=16cm]{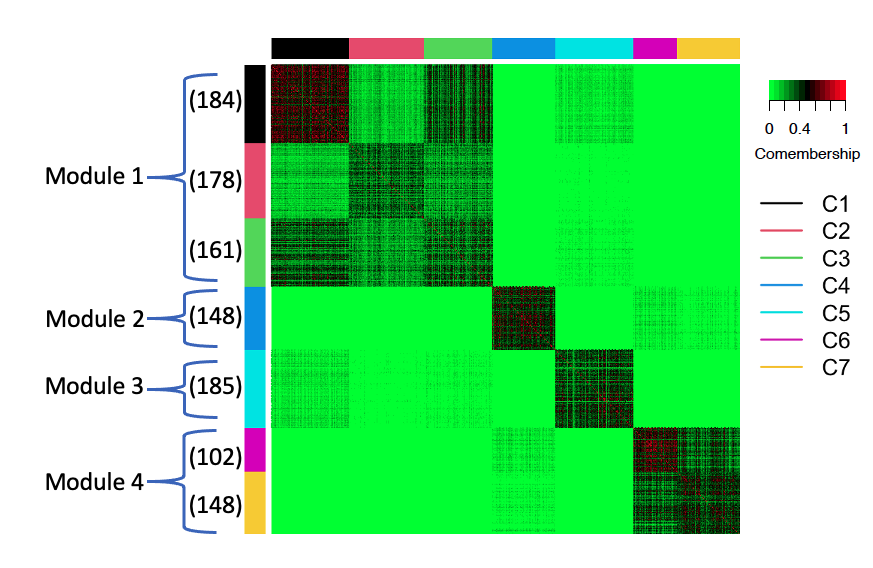}
	
	\label{fig:cluster}
\end{figure}

\begin{figure}[!thbp]
	\caption{The left panel shows the heatmap of gene expression for each cluster, along with the value of phenotypes for each sample (red color indicates higher expression and green indicates lower expression). The middle panel shows the varibility index of each gene (black indicates low varibility and white indicates high varibility). The right panel shows the weight estimation of each gene (blue represent 1, black represents 0 and yellow represents -1).}
	\vspace{2em}
	\includegraphics[width=16cm,height=16cm]{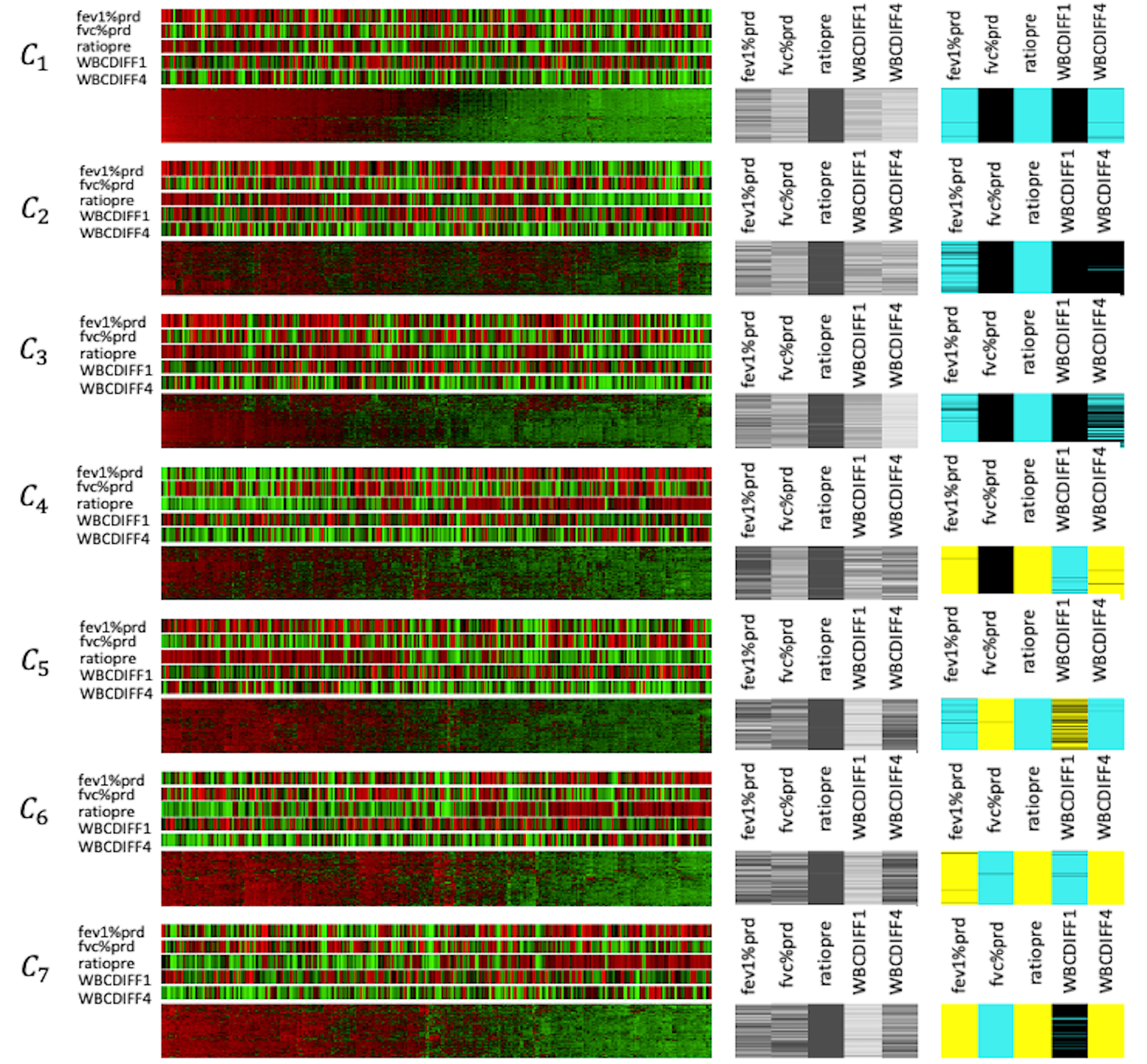}
	\label{fig:heatmap}
\end{figure}

\begin{figure}[!thpb]
	\caption{The heatmap shows the $-log_{10}(p_{jk})\times sign(\theta_{jk})$ for $j$th gene and $k$th phenotype among the genes identified in tight clustering. We truncate $-log_{10}(p_{jk})\times sign(\theta_{jk})$ to $[-10, 10]$ for better visualization. Green (-10) means a gene is negatively associated with the phenotype and red (10) means a gene is positively associated with the phenotype. }
	\includegraphics[width=15cm,height=15cm]{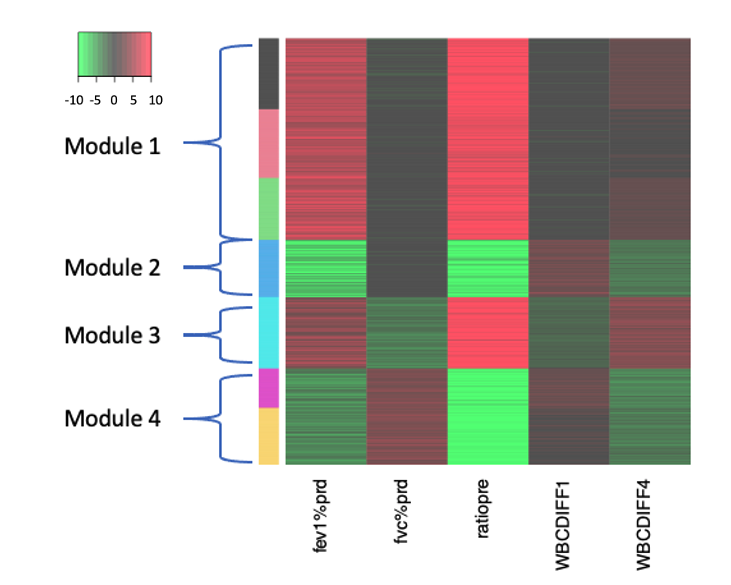}
	
	\label{fig:tight_pvalue}
\end{figure}

Following Section \ref{sec:categorization}, we calculate the comembership matrix of 4367 significant genes from AFp and utilize tight clustering algorithm to cluster genes. 1106 genes are clustered into seven clusters (C1, C2 $\cdots$ C7), where C1, C2 and C3 are more closer to one another compared with other clusters, and categorized as module 1 (Figure \ref{fig:cluster}). Similarly, C6 and C7 are combined as module 4. The left panel in Figure \ref{fig:heatmap} shows the heatmap for all the 1106 genes identified by tight clustering algorithm along with phenotype values, the middle panel indicates the varibility index for each gene and the right panel shows the weight estimation. It again indicates that C1, C2 and C3 have similar pattern (up-regulate fev1\%prd and ratiopre and no association with fvc\%prd and WBCDIFF1) and C6 and C7 have similar patten (down-regulate fev1\%prd, ratiopre and WBCDIFF4 and up-regulate fvc\%prd and WBCDIFF1), which is also confirmed by Figure \ref{fig:tight_pvalue}, the heatmap of directed -log10 (p-value) of 1106 genes selected by tight clustering method. 

	

\begin{table}[!thpb]
	
	\caption{\hspace{10pt}The proportion of weight estimated to be 1 for significant genes (4367 and 3287 for AFp and AFz respectively determined by bonferroni correction with cutoff 0.05) of AFp and AFz methods.}
	\label{table:weight}
	\begin{tabular}{|l|l|l|l|l|l|}
		\hline
		Method & fev1\%prd & fvc\%prd & ratiopre & WBCDIFF1 & WBCDIFF4 \\ \hline
		AFp    & 79\%      & 48\%     & 99\%     & 29\%     & 67\%     \\ \hline
		AFz    & 14\%      & 8\%      & 98\%     & 1\%      & 6\%      \\ \hline
	\end{tabular}
\end{table}

We next conduct pathway enrichment analysis using Fisher's exact test based on the Gene Ontology (GO), KEGG and Reactome pathway databases to assess the biological relevance of genes and show top 10 significant pathway for each module (Table \ref{table:AW_pathway}). The top pathways for different modules depict distinct aspects of lung diseases. The top pathways in module 1 involves many DNA damage \citep{adcock2018breaking,sears2019dna} and amino acid alternation/degradation pathways \citep{engelen2003altered, ubhi2012targeted}, which are known to be related to COPD in the literature. Module 2 enriches in many immune response pathways. The immune system needs to react promptly and adequately to potential dangers posed by these microbes and particles, while at the same time avoiding extensive tissue damage and many studies have shown the association between immune response and lung diseases, such as Toll-like receptor and NOD-like receptor \citep{chaput2013nod,sarir2008cells} and kinase-based protein signaling cascades \citep{mercer2006emerging}. Module 3 clearly indicates many extracellular structure pathways which provide structural support and stability to the lung. Changes in the ECM in the airway or parenchymal tissues are now recognized in the pathological profiles of many respiratory diseases including COPD \citep{burgess2016extracellular}. The top pathways in module 4 includes pathways related to cancer and vasculature development. COPD is a risk factor for lung cancer and they have many shared driving factor and genetic effect \citep{durham2015relationship}. Also, COPD is a risk factor for major cancers developing outside of the lung, including bladder cancer and pancreatic cancer \citep{divo2012comorbidities,ahn2020cancer}. Furthermore, Angiogenesis (vasculature development) is a shared phenomenon for both cancer and COPD \citep{matarese2012angiogenesis}, which may indicate the molecular connection between COPD and cancers.



\section{Discussion}
\label{sec:discussion}
In this paper, we extend AFp and AFz methods to the scenario of correlated phenotypes based on combinatorial searching for the optimal weight and permutation for determining the significance. Compared with traditional methods targeting at UIT test between each gene and all phenotypes, AFp and AFz can determine the heterogeneity among phenotypes. followed by a bootstrapping algorithm to calculate varibility index and categorize genes. From extensive simulations and the real application, we clearly show that AFp and AFz have robust performance in terms of statistical power under all scenarios. Moreover, AFp has better sensitivity of weight estimation compare with AFz, especially when one phenotype, compared with the others, has much stronger assocation with the gene. AFz tends to assign weight 1 only to the phenotype with strongest association and give weight 0 to all the other phenotypes, which forbids the further gene categorization. In summary, AFp is the method we recommend with superior performance in statistical power, determination of phenotype heterogeneity, gene categorization and biological interpretation by pathway enrichment analysis.

There are three potential limitations in the current study. Firstly, AFp needs permutation to calculate the null distribution and bootstrapping to obtain the comembership matrix for gene categorization which may need heavy computing. To relieve computational burden, we utilize R package ``Rfast" \citep{papadakis2017package} to speed up and also optimize our code to make it in an affordable range for general omics applications. To benchmark computing time, lung disease application ($K=5$, $N=279$ and $p=15,966$) with 50 times bootstrapping using 50 computing threads takes approximately 2 hours to implement AFp method. Secondly, the categorization of genes involves clustering the comembership matrix. Due to the reason that many genes may scatter around and only a small percentage of genes can generate tight clusters, we applied tight clustering algorithm to remove noise genes and generate very tight and striking clusters (Figure \ref{fig:cluster}). However, many other clustering algorithms (hierarchical clustering, $K$-means, self-organizing maps (SOM)) may be worthwhile to try. Thirdly, since AFp uses combinatorial searching for optimal weight, the number of phenotypes should be reasonable. The number of phenotypes is 10 and 5 respectively in simulation and real application parts of this paper and our software can handle this scale well. We would say that in a real application with $5000 \sim 20,000$ genes and $200 \sim 1000$ samples, usually the number of phenotypes is recommended to be below 10 for computational consideration. Users are suggested to pre-screen phenotypes and only use the phenotypes with biological insight for AFp model.

An R package to implement AFp is available on \url{https://github.com/YujiaLi1994/AFp}, along with all data and source code used in this paper. 

\begin{table}[!thbp]
	\caption{\hspace{10pt}The pathway enrichment analysis of each module by GO, KEGG and Reactome pathway database. The * sign indicates the P-value is significant under False discovery rate 0.05.}
	\label{table:AW_pathway}
	\small
	\resizebox{\textwidth}{!}{
		\begin{tabular}{|l|l|}
			\hline
			\multicolumn{1}{|c|}{pathway}                                                            & pvalue   \\ \hline
			\multicolumn{2}{|c|}{module 1}                                                                      \\ \hline
			GO:BP double-strand break repair                                                         & 4.37e-03 \\ \hline
			Reactome Double-Strand Break Repair                                                      & 4.37e-03 \\ \hline
			KEGG Valine, leucine and isoleucine degradation                                          & 6.78e-03 \\ \hline
			Reactome Branched-chain amino acid catabolism                                            & 7.82e-03 \\ \hline
			Reactome Homologous recombination repair of replication-independent double-strand breaks & 1.42e-02 \\ \hline
			GO:MF phosphotransferase activity, phosphate group as acceptor                           & 1.98e-02 \\ \hline
			GO:BP gamete generation                                                                  & 2.67e-02 \\ \hline
			GO:BP sexual reproduction                                                                & 2.68e-02 \\ \hline
			GO:MF motor activity                                                                     & 3.39e-02 \\ \hline
			GO:MF nucleobase-containing compound kinase activity                                     & 3.39e-02 \\ \hline
			\multicolumn{2}{|c|}{module 2}                                                                      \\ \hline
			KEGG Toll-like receptor signaling pathway                                             & 8.47e-06* \\ \hline
			KEGG NOD-like receptor signaling pathway                                            & 1.03e-05* \\ \hline
			KEGG MAPK signaling pathway                                                          & 3.91e-05* \\ \hline
			GO:BP response to stress                                                              & 3.95e-05* \\ \hline
			KEGG Cytosolic DNA-sensing pathway                                                    & 4.13e-05* \\ \hline
			GO:MF enzyme binding                                                                  & 7.46e-05* \\ \hline
			GO:BP protein kinase cascade                                                         & 8.41e-05* \\ \hline
			GO:MF rho gtpase activator activity                                            & 8.93e-05* \\ \hline
			Reactome NFkB and MAP kinases activation mediated by TLR4 signaling repertoire        & 1.28e-04* \\ \hline
			GO:BP regulation of protein kinase activity                                          & 1.55e-04* \\ \hline
			\multicolumn{2}{|c|}{module 3}                                                                      \\ \hline
			Reactome Extracellular matrix organization                                            & 1.48e-08* \\ \hline
			Reactome Collagen formation                                                           & 1.64e-06* \\ \hline
			GO:CC proteinaceous extracellular matrix                                             & 3.28e-06* \\ \hline
			GO:CC extracellular matrix                                                           & 3.97e-06* \\ \hline
			GO:CC extracellular region part                                                  & 2.66e-05* \\ \hline
			GO:CC collagen trimer                                                               & 7.31e-05* \\ \hline
			GO:CC extracellular region                                                           & 9.26e-05* \\ \hline
			GO:CC extracellular matrix component                                                  & 1.33e-04* \\ \hline
			GO:MF glycosaminoglycan binding                                                          & 3.30e-04 \\ \hline
			Reactome Diabetes pathways                                                               & 3.63e-04 \\ \hline
			\multicolumn{2}{|c|}{module 4}                                                                      \\ \hline
			KEGG MAPK signaling pathway                                                              & 1.87e-04 \\ \hline
			KEGG Dorso-ventral axis formation                                                        & 1.89e-04 \\ \hline
			KEGG Bladder cancer                                                                      & 2.62e-04 \\ \hline
			KEGG Pancreatic cancer                                                                   & 2.84e-04 \\ \hline
			GO:MF neurotransmitter binding                                                           & 3.80e-04 \\ \hline
			GO:BP angiogenesis                                                                       & 4.28e-04 \\ \hline
			KEGG Pathways in cancer                                                                  & 4.55e-04 \\ \hline
			GO:BP organ development                                                                  & 5.51e-04 \\ \hline
			GO:BP vasculature development                                                            & 7.34e-04 \\ \hline
			GO:BP anatomical structure formation involved in morphogenesis                           & 9.83e-04 \\ \hline
		\end{tabular}
	}
\end{table}

\section*{Acknowledgements}
YL, YF and GCT are supported by NIH R01CA190766 and R21LM012752.

\bibliographystyle{biom} 
\bibliography{ref}

\end{document}




\pagerange{\pageref{firstpage}--\pageref{lastpage}} 
\volume{63}
\pubyear{2007}
\artmonth{December}

\doi{10.1111/j.1541-0420.2005.00454.x}

\label{firstpage}
\maketitle

\section*{Web Appendix A: Details of existing methods of selecting $K$ for $K$-means}
\section*{Estimation by Cluster Tightness}
As shown in Table 1 in the main manuscript, many classical methods for determining $K$ are based on cluster tightness using within-cluster dispersion $W_K=\sum\limits_{j=1}^p WCSS_j (C_K)$, where $C_K$ is the output clustering result given $K$. The within-cluster dispersion $W_K$ is a decreasing function with respect to $K$ and the underlying true $K$ is usually reflected as an elbow point. Specifically, $W_K$ initially drops quickly and the decrease flattens markedly after the underlying true $K$ (See Web Figure 1). Detection of such an elbow point in real data is often subjective and difficult. Many estimation methods depend on an index or transformation of $W_K$ to amplify the signal and capture the elbow point by optimization or by a certain decision rule. For example, \cite{calinski1974dendrite} proposed CH index to select $K$ to maximize $\frac{BCSS(k)/(k-1)}{WCSS(k)/(n-k)}$, where $BCSS$ is the between-cluster sum of squares ($BCSS(k)=TSS-WCSS(k)$) and $TSS$ is the total sum of squares. \cite{milligan1985examination} performed a comprehensive comparison of 30 variety of indexes and concluded that the CH index was one of the best performers. \cite{krzanowski1988criterion} proposed a KL index by maximizing $\frac{DIFF(k)}{DIFF(k+1)}$, where $DIFF(k)=(k-1)^{2/p}W_{k-1}-k^{2/p}W_k$ and $p$ is the number of features. \cite{hartigan1975clustering} proposed H index by calculating $H(k)=(\frac{W_k}{W_{(k+1)}}-1)\times(n-k-1)$ and then $K$ is estimated as the smallest $k$ such that $H(k)\leq 10$. \cite{rousseeuw1987silhouettes} developed silhouette index by maximizing $\frac{b(i)-a(i)}{max\{a(i),b(i)\}}$, where a($i$) is the average dissimilarity between subject $i$ and all other subjects in the cluster to which subject $i$ belongs and b($i$) is the smallest average dissimilarity of $i$ to all points in any other cluster, of which $i$ is not a member. \cite{sugar2003finding}, based on information theoretic perspective, later proposed to maximize a jump statistic by $jump(K)=(W_K)^y-(W_{K-1})^y$ where the transformation power $y$ is typically chosen as $-p/2$ and $p$ is the total number of features. As shown in Table 1, we include all five methods, CH index, KL index, H index, silhouette and Jump, as representative summary index methods into our evaluation.

In addition to methods based on summary indexes, \cite{tibshirani2001estimating} proposed to maximize a Gap statistic defined as the difference between the observed $W_K$ and the null (reference) $W_K$ obtained from permutation where data do not contain any cluster structure. Specifically, the Gap statistic is $Gap(K)= (\sum_{b=1}^B \log(W_K^{(b)}))/B-\log(W_K)$, where $W_K^{(b)}$ is the simulated null $W_K$ from uniform distribution or PCA rotation in the $b$-th simulation and $B$ is the total number of simulations. Conceptually, subtracting the null $W_K$ from the observed $W_K$ serves to de-trend (or normalize) the decreasing pattern so that the true $K$ can be obtained by choosing the smallest $K$ such that $Gap(K)\geq Gap(K+1)-s_{K+1}$, where $s_{K+1}$ is the estimated standard deviation of $Gap(K+1)$. We will evaluate both versions of Gap statistic using uniform or PCA null reference.
\section*{Estimation by Resampling Evaluation}

Another category of methods to estimate $K$ is by resampling evaluation, including subsampling or bootstrapping. With data perturbations introduced from resampling, clustering from different resampled data should generate stable (or similar) results when the underlying true $K$ is selected.  \cite{levine2001resampling} proposed to measure the concordance between subsampled data and the original whole data to assess stability. \cite{ben2001stability} measured the stability across subsampled data and used the transition of distribution of similarity score to determine the optimal $K$. \cite{fang2012selection} compared pairwise bootstrapped data to examine the stability. We note that the Ben-Hur method is not completely quantitative since users need to manually check the transition of the distribution. Therefore, we choose the LD and FW methods as representatives of stability-based methods for comparisons. Our proposed S4 method also belongs to this category and is introduced in Section 3.

In contrast to stability-based methods, another class of methods split the original data into two portions, pretend the first portion as training data and the second portion as testing data, and mimic supervised machine learning setting to evaluate prediction accuracy. The underlying true $K$ should generate the highest prediction accuracy. Take \cite{tibshirani2005cluster} as an example, the method randomly splits data $X$ into training data $X_{tr}$ and testing data $X_{te}$. Training data are clustered into $K$ clusters (denoted as $C(X_{tr}, K)$), and the resulting $K$ cluster centroids are used as a classifier to assign test samples into $K$ clusters. The element ($i_1$, $i_2$) of co-membership matrix $D[C(X_{tr},k),X_{te}]_{i_1,i_2}=1$ if sample $i_1$ and $i_2$ of testing data are predicted in the same cluster by the training data centroids and 0 otherwise. By comparing  clustering results between testing data on training centroids and testing data on test clusters ($A_{k1},A_{k2},\cdots,A_{kk}$) for a given number of cluster $k$, the prediction strength for given $k$ is defined as: $ps(k)=\min\limits_{1\leq j \leq k}\frac{1}{n_{kj}(n_{kj}-1)}\sum\limits_{i_1,i_2 \in A_{kj}} D[C(X_{tr},k),X_{te}]_{i_1,i_2}$,
where $n_{k1},n_{k2},\cdots n_{kk}$ are the number of samples in clusters $A_{k1},A_{k2},\cdots,A_{kk}$. \cite{tibshirani2005cluster} suggests to choose the largest number of cluster $k$ with $ps(k)$ larger than a certain threshold. \cite{dudoit2002prediction} proposed clest method which uses reference data to adjust the prediction score. However, this method has been criticized to contain many unspecified parameters and hard to implement in practice \citep{lange2004stability}. \cite{lange2004stability} proposed a different framework to adjust prediction score by reference data. However, the method requires heavy computation to measure prediction score for both original and repeatedly simulated reference data and no software package was provided for implementation. Thus, only prediction strength (PS) from this category is selected for comparison in this paper.

\section*{Web Appendix B: Settings of Simulation I, II and III}
\section*{Simulation I}
\textit{\underline{Well-separated}:}
\begin{itemize}
	\item Setting A1: $K=3$ clusters in two dimensions are generated by standard normal distribution centered at (0, 0), (0, 5) and (5, 3) respectively, with 25, 25 and 50 samples in each cluster.
	\item Setting A2: $K=4$ clusters in $p=3$ dimensions are generated with centers randomly obtained by N(0, 5$\cdot I$). We use standard normal to randomly generate 25 or 50 observations for each cluster. If points of any two clusters have distance smaller than 1, we will discard this simulation and simulate the data again.
	\item Setting A3: Similar to Setting A2 but dimensionality increases to $p=5$ and the centers are randomly obtained from N(0, 4$\cdot I$). 
	\item Setting A4: Similar to Setting A2 but dimensionality increases to $p=8$ and the centers are randomly obtained from N(0, 3$\cdot I$). 
	\item Setting A5: Similar to Setting A2 but dimensionality increases to $p=10$ and the centers are randomly obtained from N(0, 1.9$\cdot I$). 
	\item Setting A6: We simulate $K=2$ clusters in $p=3$ dimensions with 100 observations in each cluster. For the first cluster, choose $x_1=x_2=x_3=t$ where t is chosen by equal spaced values from -0.5 to 0.5, then add Gaussian  noise with standard deviation 0.1 to each feature. The second cluster are generated in the same way except for adding value 10 to each feature at the end. This forms two elongated cluster on main diagonal in three-dimensional cube.
	\item Setting A7: Similar to Setting A6, but instead of adding value 10, we only add value 1 to each feature in the second cluster, producing two close and elongated clusters.
\end{itemize}

\textit{\underline{Non-well-separated}:}
\begin{itemize}
	\item Setting B1: There are $K=4$ clusters in two dimensions and each cluster is generated from standard normal distribution centered at (0,0), (0, 2.5), (2.5, 0), (2.5, 2.5), with 25 observations respectively.
\item Setting B2: Similar to Setting B1 but the clusters are centered at (0,0), (0, 3), (3, 0), (3, 3).
\item Setting B3: Similar to Setting B1 but the clusters are centered at (0, 0), (0, 3.5), (3.5, 0), (3.5, 3.5).
\item Setting B4: We simulate $K=2$ clusters in $p=5$ dimensions with 50 observations in each cluster. All the features are generated from standard norm and then a constant shift 2 is added to the first feature of one cluster.
\item Setting B5: We simulate $K=2$ clusters in $p=10$ dimensions with 50 observations in each cluster. All the features are generated from standard norm and then a constant shift 2 is added to the first feature of one cluster.
\end{itemize}

Setting A1, A2, A5 and A6 are replication of simulation studies presented by \cite{tibshirani2001estimating}. Setting A3 and A4 are modified from Setting A2 and Setting A7 is modified from A6. Setting B1-B5 are non-well-separated simulation settings in terms of lower adjusted rand index (See Table 2 in the main manuscript) and difficulty to separate them from Null data ($K=1$, See Section 7). All the simulation settings are repeated for 100 times and the searching space of number of clusters is chosen from 2 to 10. 

\section*{Simulation II}
The following simulation is designed to evaluate methods (Gap statistic, prediction strength and S4) for determining $K$ and $\lambda$ in sparse $K$-means under independent feature scenario. The implementation of Gap statistic is based on the sparcl R package and the reference data are generated by permutation in the package. We simulated three clusters, each with 33 subjects, and each subject has 1,000 features, of which $q$ features are informative to distinguish the three clusters while other features are random noises. Denote by $ X_{n\times p}$ the data matrix where $n=99$ and $p=1,000$ and $x_{i,1:j}$ is the vector of subject $i$ with features from 1 to $j$. We simulate features by multivariate normal distribution. For the first $q$ predictive features, $x_{i,1:q} \sim mvrnorm(u, I_q)$ for $1 \leq i \leq 33$, $x_{i,1:q} \sim mvrnorm(0, I_q)$ for $34 \leq i \leq 66$, and $x_{i,1:q} \sim mvrnorm(-u, I_q)$ for $67 \leq i \leq 99$, where $u$ is the effect size to distinguish three clusters. For the remaining noise features, $x_{i,(q+1):p} \sim mvrnorm(0, I_{p-q})$ for $1\leq i\leq 99$. We choose $q=(50,200)$ and $u=(0.4, 0.6, 0.8)$ to generate six settings and repeat each setting for 50 times. We perform $B=100$ resampling evaluation for all three methods and choose number of cluster $K$ from 2 to 7 for all six simulation settings. 

Next, we evaluate the methods under two situations. Firstly, we assume $K=3$ is known and compare the performance of estimating $\lambda$, the sparsity parameter, and this is the same setting considered in \cite{witten2010framework}. Secondly, we consider simultaneous estimation of $K$ and $\lambda$. For both situations, we benchmark the clustering accuracy by adjusted Rand index (ARI) \citep{hubert1985comparing} when comparing to the underlying true clustering structure. We also benchmark feature selection by comparing selected features to the underlying true predictive features using Jaccard index \citep{jaccard1901distribution}, defined as $J(A,B)=A\cap B/A\cup B$ where $A$ is the set of selected features from sparse $K$-means and $B$ represents the set of $q$ true features. Root mean square error (RMSE) of $K$ estimation is used to evaluate the performance of estimating $K$, when simultaneous estimation of $K$ and $\lambda$ is considered.

\section*{Simulation III}

To better mimic the nature of gene expression profile data from microarray or RNA-seq experiments, a typical high-dimensional data type for clustering, we simulate data of three clusters with gene correlation structure as co-regulated gene modules. The purpose is usually to cluster patients to identify novel disease subtypes in such applications. We construct six simulation settings with different effect size $(U_{upper},U_{lower})$ and covariance structure within gene modules ($\phi_{cov}$). Below are the detailed steps to simulate cluster predictive genes with different covariance structure and effect size, as well as noise genes.

\textit{\underline{Simulation of cluster predictive genes}:} 
\begin{itemize*}
	\item[1.]Simulate the number of subjects $N_1$, $N_2$ and $N_3$ for three disease subtypes by sampling from Poisson distribution with mean 40, 30 and 20 respectively. The total number of subjects in each simulated data is $N=N_1+N_2+N_3$
	\item[2.]Simulate $M$ gene modules. In each module, sample $n_m(1\leq m\leq M)$ genes from $POI(20)$. Therefore, there will be an average of $20\times M$ predictive genes to characterize the three clusters (disease subtypes).
	\item[3.]Simulate $u_{km}\sim U(4, 10)$ with constrain $U_{lower} \leq max_{p,q}u_{pm}-u_{qm} \leq U_{upper}$, where $u_{km}$ is the template gene expression of cluster $k$ $(1 \leq k \leq 3)$ and module $m$ $(1\leq m \leq M)$ and $(U_{lower},U_{upper})$ reflects effect size.
	\item[4.]Add biological variation $\sigma_1^2$ to the template gene expression and simulate $X_{kmi}\sim N(u_{km},\sigma_1^2)$ for each module $m$, subject $i$ ($1\leq i \leq N_k$) of cluster $k$.
	\item[5.]Simulate covariance matrix $\Sigma_{mk}$ for genes in module $m$ ($1\leq m \leq M$) and cluster $k$ ($1\leq k \leq 3$). First simulate $\Sigma^{'}_{mk}$ from inverse Wishart distribution, $W^{-1}(\Phi,60)$ where $\Phi=(1-\phi_{cov})\cdot I_{n_m\times n_m}+\phi_{cov}\cdot J_{n_m\times n_m}$, $I$ is identity matrix, $J$ is a matrix with all elements equivalent to 1 and $\phi_{cov}$ is a scalar controlling degree of correlation among genes, Then $\Sigma_{mk}$ is calculated by standardizing $\Sigma^{'}_{mk}$ such that the diagonal elements are all 1.
	\item[6.]Simulate gene expression levels of genes in module $m$ for sample $i$ in cluster $k$ as $(X_{1kmi}, \cdots, X_{n_mkmi})$ from multivariate normal distribution with mean $X_{kmi}$ and covariance matrix $ \Sigma_{mk}$, where $1\leq i\leq N_k$, $1 \leq m \leq M$, $1\leq k \leq 3$.
\end{itemize*}

\textit{\underline{Simulation of noise genes}:} 
\begin{itemize}
	\item[1.]Simulate 600 noise genes.  For each gene, we generate the mean template gene expression $u_g \sim U(4, 10)$, where $1\leq g \leq 600$. 
	\item[2.]Then we add biological variation variance $\sigma_2^2$ to simulate gene expression level $X_{gi} \sim N(u_g, \sigma^2_2)$, $1 \leq i \leq N$. 
\end{itemize}

We fix $\sigma_1^2=0.2, \sigma_2^2=1, M=10$ while tuning effect size $(U_{upper},U_{lower})$ and correlation parameter $\phi_{cov}$ to compare S4 with Gap and PS in different scenarios. Since the number of predictive genes in each gene module follows $POI(20)$, so the average number of predictive genes in each dataset is 200. Number of clusters $K$ is selected from 2 to 7 and ARI, Jaccard and RMSE of $K$ are used to compare the performance. Each setting is repeated 50 times.

\section*{Web Appendix C: Compare Two-stage method and one-stage method for simultaneously estimating $K$  and $\lambda$}	
After the clustering $S^*_{\rho}(K, \lambda)$ and feature selection $F(K, \lambda)$ concordance scores are defined, a naive one-stage approach to estimate $K$ and $\lambda$ is by optimizing sum of the two concordance scores: $(\hat{K},\hat{\lambda})= \arg \max\limits_{K, \lambda} S^*_{\rho}(K,\lambda)+F(K,\lambda)$.  However, the approach we propose in this paper is a two-stage approach which first obtain $\hat{K}$ by $\hat{K}= \arg\max\limits_{K} \left(\max\limits_{\lambda} S^*_{\rho}(K,\lambda)\right)$.
Next, given $\hat{K}$, we estimate $\hat{\lambda}$ by $\hat{\lambda}= \arg \max\limits_{\lambda} S^*_{\rho}(\hat{K},\lambda)+F(\hat{K},\lambda)$.
The simulation below illustrate why two-stage approach has advantages over the one-stage one.

We here simulate data of three clusters where two clusters are closer to each other compared to the third one. Denote the whole data matrix by $ X_{n\times p}$ where $n=99$ and $p=300$ and $x_{i,1:j}$ is a vector of subject $i$ with feature from 1 to $j$. We simulate features by multivariate normal distribution.
\begin{itemize}
	\item Simulate cluster 1: for $1 \leq i \leq 33,$ $x_{i,1:50} \sim mvrnorm(3, I_{50})$, $x_{i,51:150} \sim mvrnorm(0.6, I_{100})$, $x_{i,151:300} \sim mvrnorm(0, I_{150})$.
	\item Simulate cluster 2: for $1 \leq i \leq 33,$ $x_{i,1:50} \sim mvrnorm(-1, I_{50})$, $x_{i,51:150} \sim mvrnorm(0, I_{100})$, $x_{i,151:300} \sim mvrnorm(0, I_{150})$.
	\item Simulate cluster 3: for $1 \leq i \leq 33,$ $x_{i,1:50} \sim mvrnorm(-1, I_{50})$, $x_{i,51:150} \sim mvrnorm(-1.5, I_{100})$, $x_{i,151:300} \sim mvrnorm(0, I_{150})$.
\end{itemize}	

Number of cluster $K$ is chosen from 2 to 7.  We do simulation for 50 times and each time we generate $B=100$ subsample. The results show that two-stage approach always chooses $K^*=3$ whereas one-stage approach always chooses $K^*=2$. In this simulation setting, feature $1\sim50$ can well separate cluster 1 from the other two clusters and  feature $51\sim150$ can separate all three clusters. Therefore, $S^*_{\rho}(K=2,\lambda)$ and  $S^*_{\rho}(K=3,\lambda)$  will both be 1 if $\lambda$ is large enough. However, $F(K=3, \lambda)$ is lower than $F(K=2, \lambda)$ since feature $1\sim50$ and feature $51\sim150$ both contribute to the clustering if $K=3$ and feature selection is unstable compared with $K=2$ where only the feature $1\sim50$ strongly contribute to the clustering. Therefore, since one-stage approach chooses $K^*$ and $\lambda^*$ by $\arg \max\limits_{K, \lambda} S^*_{\rho}(K,\lambda)+F(K,\lambda)$, it chooses $K^{*}=2$. The two-stage approach chooses $K^*=3$ since it only uses $S^*_{\rho}(K,\lambda)$ when estimating $K$.

\section*{Web Appendix D: Extended Prediction Strength for Estimating $K$ and $\lambda$}
We also extend prediction strength method \citep{tibshirani2005cluster} as $ps(k,\lambda)=\min\limits_{1\leq j \leq k}\frac{1}{n_{kj,\lambda }(n_{kj,\lambda}-1)}\sum\limits_{i_1,i_2 \in A_{kj,\lambda }} D[C(X_{tr},k),X_{te}]_{i_1,i_2}$ to simultaneously estimate $K$ and $\lambda$ for sparse $K$-means. All the notation are similar to those of prediction strength in Web Appendix A. The only difference lies in $D[C(X_{tr},k),X_{te}]_{i_1,i_2}$. When using training data centroid to predict test samples, instead of using Euclidean distance, here we use weighted Euclidean distance and the weights are obtained by the result of sparse $K$-means of the training data. In addition, we use the features selected by training data to compare predictive features of testing data and measure feature concordance score. Following similar rationale of S4, denote by $f^{(tr)}_j$ as the feature selection index for feature $j$ of the training data (i.e., $f^{(tr)}_j=1$ if feature $j$ is selected otherwise $f^{(tr)}_j=0$). Similarly, define $f_j^{(te)}$ the feature selection index for feature $j$ from test data. We define the feature prediction strength as $F_{ps}(K, \lambda)=\frac{\sum _{j=1}^p f^{(tr)}_{j}I\{f_{j}^{(te)}=1 \}}{\sum_{j=1}^p I\{f_{j}^{(te)}=1 \}}$. Using the similar idea of prediction strength by \cite{tibshirani2005cluster}, we define $ps^*(K)=\max\limits_{\lambda}  ps(K,\lambda)$ and select $\hat{K}$ as the largest $K$ with $ps^*(K)\geq s_0$, where $s_0=0.8$ is the default threshold. If no $ps^*(K)$ is larger than $s_0$, $\hat{K}=\max\limits_{K} ps(K) $. Given $\hat{K}$, $\hat{\lambda}= \arg \max\limits_{\lambda} ps(\hat{K},\lambda)+F_{ps}(\hat{K},\lambda)$, similar to S4.

\section*{Web Appendix E: Sensitivity analysis for trimmed mean and number of subsampling in S4}
As described in the toy examples in Web Figures 3-5, adequate trimming by cluster concordance score before averaging could exclude scattered points and improve estimation performance. In Web Table 1, we perform a sensitivity analysis of different trimming parameter $\rho=0, 2, 5, 8, 10, 15, 20$ by applying S4 to the 12 simulation settings in simulation I. We find that $\rho=5$ and $\rho=8$ works reasonably well in all simulations. To be conservative, throughout this paper, we set $\rho=5$ in all comparisons. For the number of subsampling $B$, sufficiently large $B$ is desired to provide a stable result while it demands more computing. In Web Table 2, we perform a sensitivity analysis of different number of subsampling $B=20, 50 ,100, 200, 500$ and find that for some simulations (i.e., A1-A5, B1, B4 and B5), $B$=20 is already stable enough. For A6, A7, B2 and B3, a larger $B$ is needed to stabilize the result. Since $B=100$ generally generates stable results, we set $B=100$ for all simulations.  For the applications, we use $B=500$ to further stabilize the performance.

\section*{Web Appendix F: Data description of nine real applications}
\label{describe}
\subsection*{Microarray  datasets}
\textit{\underline{Three leukemia datasets}:} Following \cite{huo2016meta}, we collect three leukemia transcriptomic studies for evaluation: \cite{verhaak2009prediction}, \cite{balgobind2011evaluation}, and \cite{kohlmann2008international}. For each study, we only consider samples from acute myeloid leukemia with $K=3$ pre-detected chromosome translocation subtypes: inv(16)(inversions in chromosome 16), t(15;17)(translocations between chromosomes 15 and 17), and t(8;21)(translocations between chromosomes 8 and 21). All the datasets are downloaded directly from NCBI GEO website with GSE6891, GSE17855, and GSE13159. From the original 54,676 probesets in each dataset, we remove the probesets with missing values and select the probesets with the largest interquartile range to represent the gene if multiple probesets are mapped to the same gene. 20,192 unique genes are remained for each study after this preprocessing. For each study, we further transform data to log scale and only keep the top 10,000 genes with the largest mean expression level (i.e. filter out low-expressed genes). We next filter out 8,000 genes with smaller variance (i.e. genes with little predictive information). Finally, the remaining $p=2,000$ genes are used in the analysis.

\noindent \textit{\underline{Mammalian tissue types dataset}:} Gene expression from human and mouse samples across a diverse array of tissues, organs,
and cell lines have been profiled by \cite{su2002large}. Here we only consider $K=4$ tissue types: breast, prostate, lung, and colon, which is available in R package fabiaData (Hochreiter et al., 2013) and website \url{http://portals.broadinstitute.org/cgi-bin/cancer/datasets.cgi}. The original dataset has $n=102$ samples and 5,565 probesets (genes). Following similar preprocessing procedure above, we keep 3,000 genes with the highest mean expression value and then $p=2,000$ genes are used in the analysis after further filtering low-variance genes.

\subsection*{ RNA sequencing data}
\textit{\underline{Multiple brain regions of rat dataset}:} \cite{li2013transcriptome} generated a rat experiment including multiple brain regions (GSE47474) . RNA samples from three brain regions (hippocampus, striatum and prefrontal cortex) were sequenced for both control strains and HIV infected strains. Only the 36 control strains (12 in each brain region) are used here to see whether samples from three brain regions can be correctly clustered ($K$ = 3; $n_1$ = $n_2$ = $n_3$ $= 12$). The original count data are transformed into CPM values followed by log transformation and then $p=2,000$ genes are kept by filtering low-expressed genes and low-variance genes.

\noindent \textit{\underline{Pan-cancer dataset}:} We download a dataset which is part of The Cancer Genome Atlas (TCGA) pan-cancer analysis project, available at the UCI machine learning repository (\url{https://archive.ics.uci.edu/ml/datasets/gene+expression+cancer+RNA-Seq#}). This collection of data consists of $K=5$ different types of tumor: 300 breast cancer (BRCA), 146 kidney clear cell carcinoma (KIRC), 78 colon cancer (COAD), 141 lung adenocarcinoma (LUAD) and 136 prostate cancer (PRAD). The dataset has already been normalized and we use the same filtering process to keep $p=2,000$ genes.\\

\subsection*{SNP dataset}
The SNP dataset was previously applied in \cite{witten2010framework}, where they showed that when number of cluster is known as three, the gap statistic will seemingly overestimate the number of features with non-zero weight. The data is publicly available from Haplotype Map (HapMap) project of the International HapMap Consortium. Following the same preprocessing procedure as \cite{witten2010framework}, only phase III SNP data is used and we restrict the analysis to chromosome 22 of $K=3$ populations: African ancestry in southwest USA (ASW), Utah residents with European ancestry (CEU), and Han Chinese from Beijing (CHB) since these three populations are known to be genetically distinct. All the available SNPs on chromesome 22 are considered in the data, which accounts for $n=293$ samples and $p=17,026$ SNPs. We then coded AA as 2, Aa as 1 and aa as 0, and use 5-nearest neighbors method \citep{troyanskaya2001missing} to impute the missing data.

\subsection*{Non-Omics data}
\noindent \textit{\underline{Plant species leaves dataset}:} \cite{mallah2013plant} introduced a dataset consisting of 100 species of plants with three types of features for leaves: shape, texture and margin. Here we only consider $K=4$ species: Acer Mono, Acer Palmatum, Acer Pictum and Acer Capillipes. After deleting features with any missing values, we have $n=64$ samples (16 for each species) and $p=187$ features.

\noindent \textit{\underline{ISOLET Data Set}:} ISOLET dataset was generated by a study where 150 subjects spoke each letter of the alphabet twice and recorded $p=617$ features including spectral coefficients, contour features, sonorant features, pre-sonorant features and post-sonorant features. We only use $K=5$ vowels and $n=1,200$ training subjects (240 samples for each of five vowels). Both plant species dataset and ISOLET dataset are publicly available in the UCI machine learning repository.

\section*{Web Figures and Tables}
\begin{figure}[!htb]
	
	\centering
	\subfigure[Three-clusters data]{                    
		\begin{minipage}[A]{0.48\textwidth}
			\centering                                                     
			\includegraphics[width=8cm,height=8cm]{fig/Supplement1a.pdf}               
	\end{minipage}}
	\subfigure[ $W_{K}$ for three-clusters data]{
		\begin{minipage}[B]{0.48\textwidth}
			\centering                                                     
			\includegraphics[width=8cm,height=8cm]{fig/Supplement1b.pdf}               
		\end{minipage}
	}
	\vfill
	\centering
	\subfigure[Two-clusters data]{                    
		\begin{minipage}{0.48\textwidth}
			\centering                                                     
			\includegraphics[width=8cm,height=8cm]{fig/Supplement1c.pdf}               
	\end{minipage}}
	\subfigure[ $W_{K}$ for two-clusters data]{
		\begin{minipage}{0.48\textwidth}
			\centering                                                     
			\includegraphics[width=8cm,height=8cm]{fig/Supplement1d.pdf}               
		\end{minipage}
	}
	\caption{Illustration of elbow point of within-cluster dispersion $W_{K}$}
	
\end{figure}

\begin{figure}[!htb]
	\centering
	\includegraphics[width=10cm,height=10cm]{fig/Figure1a.pdf} 
	\caption{Three clusters where the left two are close}
\end{figure}

\begin{figure}[!htb]
	\centering
	\includegraphics[width=16cm,height=16cm]{fig/Sim2.pdf}     
	\caption{The ordered clustering concordance score for simulation setting A1  from $K=2$ to $K=5$ (underlying truth is $K=3$), the dashed line is the 5\% timmed line.}
\end{figure}
\begin{figure}[!htb]
	\centering
	\includegraphics[width=16cm,height=16cm]{fig/Sim3.pdf}     
	\caption{The ordered clustering concordance score for simulation setting A2  from $K=2$ to $K=5$ (underlying truth is $K=4$), the dashed line is the 5\% timmed line.}
\end{figure}
\begin{figure}[!htb]
	\centering
	\includegraphics[width=16cm,height=16cm]{fig/Sim4.pdf}  
	\caption{The ordered clustering concordance score for simulation setting A5  from $K=2$ to $K=5$ (underlying truth is $K=4$), the dashed line is the 5\% timmed line.}   
\end{figure}

\begin{table}[!htp]
	\captionsetup{font=normal}
	
	\caption {Sensitivity analysis for trimmed mean in S4: Simulation result for trimming different proportions of samples in clustering without feature selection. Each cell indicates frequency of each method choosing the correct $K$ in different settings by trimming different proportion of samples.}\label{table:Trimmedmeans}
	\label{comp} 
	\hspace{-2cm}
	\begin{center}
		\begin{tabu}{cccccccccc}
			
			\hline
           \diagbox{Setting}{$\rho \%$}& 0\% & 2\% & 5\% & 8\% & 10\% & 15\% & 20\% \\
           \hline
Setting A1 ($K=3$) & 83  & 96  & 100 & 99  & 99   & 97   & 92   \\
\hline
Setting A2 ($K=4$) & 76  & 89  & 98  & 99  & 99   & 99   & 98   \\
\hline
Setting A3 ($K=4$) & 51  & 69  & 86  & 88  & 90   & 94   & 98    \\
\hline
Setting A4 ($K=4$) & 51  & 70  & 85  & 94  & 94   & 96   & 97   \\
\hline
Setting A5 ($K=4$) & 40  & 54  & 72  & 84  & 88   & 92   & 94   \\
\hline
Setting A6 ($K=2$) & 100 & 99  & 90  & 60  & 39   & 2    & 0    \\
\hline
Setting A7 ($K=2$) & 58  & 68  & 87  & 89  & 84   & 56   & 19   \\
\hline
Setting B1 ($K=4$) & 42  & 41  & 41  & 40  & 43   & 45   & 49   \\
\hline
Setting B2 ($K=4$) & 64  & 65  & 63  & 68  & 69   & 74   & 77  \\
\hline
Setting B3 ($K=4$) & 78  & 78  & 81  & 86  & 87   & 90   & 92   \\
\hline
Setting B4 ($K=2$) & 90  & 90  & 93  & 94  & 95   & 98   & 98   \\
\hline
Setting B5 ($K=2$) & 86  & 87  & 88  & 89  & 89   & 91   & 93  \\
\hline
		\end{tabu}

	\end{center}

\end{table}

\begin{table}[!htp]
	\captionsetup{font=normal}
	
	\caption {Sensitivity analysis for number of subsampling $B$ in S4: For each dataset, we calculate variance of ARI by implementing S4 100 times using $B$=20, 50, 100, 200 and 500. 50 simulation is done to each simulation setting and the average standard deviation of ARI estimation for each simulation setting is shown. }\label{table:numberofsubsampling}
	\label{comp} 
	\hspace{-2cm}
	\begin{center}
		\begin{tabu}{ccccccccc}
			\hline
           Settings &$B$=20 & $B$=50 & $B$=100 & $B$=200 & $B$=500 \\
           \hline
Setting A1 ($K=3$) &0.043 & 0.024  & 0.007   & 0.004   & 0.000   \\
\hline
Setting A2 ($K=4$)  & 0.024& 0.022   & 0.025    & 0.032    & 0.024    \\
\hline
Setting A3 ($K=4$)  & 0.033& 0.037   & 0.039    & 0.040    & 0.046    \\
\hline
Setting A4 ($K=4$)  & 0.046& 0.062   & 0.054    & 0.046    & 0.043    \\
\hline
Setting A5 ($K=4$) &  0.068& 0.066   & 0.063    & 0.058    & 0.055    \\
\hline
Setting A6 ($K=2$)  & 0.174& 0.105   & 0.066    & 0.054    & 0.059    \\
\hline
Setting A7 ($K=2$)  & 0.174& 0.131   & 0.108    & 0.090    & 0.068    \\
\hline
Setting B1 ($K=4$)  & 0.056& 0.045   & 0.038    & 0.034    & 0.027    \\
\hline
Setting B2 ($K=4$)  & 0.085& 0.066   & 0.050    & 0.037    & 0.025    \\
\hline
Setting B3 ($K=4$) &  0.094& 0.076   & 0.062    & 0.050    & 0.035    \\
\hline
Setting B4 ($K=2$) &  0.047& 0.034   & 0.023    & 0.020    & 0.017    \\
\hline
Setting B5 ($K=2$) &  0.048& 0.042   & 0.035    & 0.030    & 0.021   \\
\hline
		\end{tabu}
	\end{center}
\end{table}

\begin{table}[]
	
		\caption {Sensitivity analysis of selecting $\lambda$ by weighted sum and geometric mean using Simulation II (features are mutually independent).}\label{table:Combine_two_score_ind}
	\begin{tabular}{llllllll}
		\hline
\multirow{3}{*}{Index}              & \multirow{3}{*}{\tabincell{c}{Methods\\($\omega1, \omega2$)} }& \multicolumn{3}{l}{q=50 predictive genes} & \multicolumn{3}{l}{q=200 predictive genes} \\ \cline{3-8} 
&                          & \multicolumn{3}{c}{effect size}           & \multicolumn{3}{c}{effect size}            \\ \cline{3-8} 
&                          & u=0.4        & u=0.6        & u=0.8        & u=0.4         & u=0.6        & u=0.8        \\ \hline
		\multirow{4}{*}{\tabincell{c}{Clustering\\accuracy\\(ARI)} }                & (0.25, 0.75)               & 0.264   & 0.442  & 0.955  & 0.513   & 0.997   & 1      \\ \cline{2-8} 
		& (0.5, 0.5)                 & 0.262   & 0.442  & 0.958  & 0.517   & 0.997   & 1      \\ \cline{2-8} 
		& (0.75, 0.25)               & 0.256   & 0.442  & 0.961  & 0.517   & 0.997   & 1      \\ \cline{2-8} 
		& Geometric mean           & 0.268   & 0.442  & 0.958  & 0.517   & 0.997   & 1      \\ \hline
		\multirow{4}{*}{\tabincell{c}{feature\\selection\\(Jaccard)}}            & (0.25, 0.75)               & 0.237   & 0.784  & 0.973  & 0.588   & 0.925   & 0.989  \\ \cline{2-8} 
		& (0.5, 0.5)                 & 0.225   & 0.785  & 0.971  & 0.606   & 0.925   & 0.989  \\ \cline{2-8} 
		& (0.75, 0.25)               & 0.191   & 0.774  & 0.968  & 0.61    & 0.925   & 0.989  \\ \cline{2-8} 
		& Geometric mean           & 0.236   & 0.785  & 0.971  & 0.604   & 0.925   & 0.989  \\ \hline
		\multirow{4}{*}{\tabincell{c}{Number\\of features\\(selected)}} & (0.25, 0.75)              & 351     & 51     & 47     & 137     & 184     & 195    \\ \cline{2-8} 
		& (0.5, 0.5)              & 339     & 45     & 54     & 220     & 184     & 195    \\ \cline{2-8} 
		& (0.75, 0.25)               & 308     & 58     & 54     & 265     & 184     & 195    \\ \cline{2-8} 
		& Geometric mean           & 414     & 45     & 54     & 137     & 184     & 195    \\ \hline
		\multirow{4}{*}{RMSE of K}          & (0.25, 0.75)               & 1.581   & 1      & 0      & 0.927   & 0       & 0      \\ \cline{2-8} 
		& (0.5, 0.5)                 & 1.581   & 1      & 0      & 0.927   & 0       & 0      \\ \cline{2-8} 
		& (0.75, 0.25)               & 1.581   & 1      & 0      & 0.927   & 0       & 0      \\ \cline{2-8} 
		& Geometric mean           & 1.581   & 1      & 0      & 0.927   & 0       & 0      \\ \hline
	\end{tabular}
\end{table}

\begin{table}[]
	
	\caption {Sensitivity analysis of selecting $\lambda$ by weighted sum and geometric mean using Simulation III (features are correlated).}\label{table:Combine_two_score_cov}
	\resizebox{\textwidth}{!}{
	\begin{tabular}{llllllll}
\hline
\multirow{3}{*}{Index}              & \multirow{3}{*}{\tabincell{c}{Methods\\($\omega1, \omega2$)}} & \multicolumn{2}{c}{$\phi_{cov}$=0.1}     & \multicolumn{2}{c}{$\phi_{cov}$=0.3}     & \multicolumn{2}{c}{$\phi_{cov}$=0.5}     \\ \cline{3-8} 
&                        & \multicolumn{2}{l}{effect size $(U_{lower},U_{upper})$} & \multicolumn{2}{l}{effect size $(U_{lower},U_{upper})$} & \multicolumn{2}{l}{effect size $(U_{lower},U_{upper})$} \\ \cline{3-8} 
&                        & (0.8, 1.0)      & (1.0, 1.5)     & (1.5, 2)        & (2, 2.5)       & (1.5, 2)        & (2, 2.5)       \\ \hline
\multirow{4}{*}{\tabincell{c}{Clustering\\accuracy\\(ARI)}}                & (0.25, 0.75)           & 0.774           & 0.972          & 0.973           & 0.999          & 0.867           & 0.981          \\ \cline{2-8} 
& (0.5, 0.5)             & 0.774           & 0.978          & 0.973           & 0.999          & 0.872           & 0.981          \\ \cline{2-8} 
& (0.75, 0.25)           & 0.775           & 0.978          & 0.973           & 0.999          & 0.873           & 0.981          \\ \cline{2-8} 
& Geometric mean         & 0.774           & 0.978          & 0.973           & 0.999          & 0.872           & 0.981          \\ \hline
\multirow{4}{*}{\tabincell{c}{feature\\selection\\(Jaccard)}}            & (0.25, 0.75)           & 0.659           & 0.874          & 0.919           & 0.988          & 0.81            & 0.931          \\ \cline{2-8} 
& (0.5, 0.5)             & 0.659           & 0.882          & 0.919           & 0.988          & 0.819           & 0.931          \\ \cline{2-8} 
& (0.75, 0.25)           & 0.665           & 0.884          & 0.919           & 0.988          & 0.832           & 0.931          \\ \cline{2-8} 
& Geometric mean         & 0.659           & 0.882          & 0.919           & 0.988          & 0.82            & 0.931          \\ \hline
\multirow{4}{*}{\tabincell{c}{Number\\of features\\(selected)}} & (0.25, 0.75)           & 85              & 164            & 157             & 169            & 50              & 126            \\ \cline{2-8} 
& (0.5, 0.5)             & 85              & 164            & 157             & 169            & 50              & 126            \\ \cline{2-8} 
& (0.75, 0.25)           & 112             & 164            & 157             & 169            & 50              & 126            \\ \cline{2-8} 
& Geometric mean         & 85              & 164            & 157             & 169            & 50              & 126            \\ \hline
\multirow{4}{*}{RMSE of K}          & (0.25, 0.75)           & 0.693           & 0.2            & 0.245           & 0              & 0.51            & 0.2            \\ \cline{2-8} 
& (0.5, 0.5)             & 0.693           & 0.2            & 0.245           & 0              & 0.51            & 0.2            \\ \cline{2-8} 
& (0.75, 0.25)           & 0.693           & 0.2            & 0.245           & 0              & 0.51            & 0.2            \\ \cline{2-8} 
& Geometric mean         & 0.693           & 0.2            & 0.245           & 0              & 0.51            & 0.2    \\
\hline
	\end{tabular}
}
\end{table}

\begin{table}[]
	\caption{Clustering without feature selection simulation result: determining $K$ for $K$-means. Every row indicates the average of $K$ estimated from each method.}\label{table:LowdimensionsimulationARI}
	\begin{tabular}{llllllllllllll}
		\hline
		Category                                                                     & Method     & \rotatebox{90}{A1 ($K$=3)}   & \rotatebox{90}{A2 ($K$=4)}   & \rotatebox{90}{A3 ($K$=4)}   & \rotatebox{90}{A4 ($K$=4)}   & \rotatebox{90}{A5 ($K$=4)}   & \rotatebox{90}{A6 ($K$=2)}  & \rotatebox{90}{A7 ($K$=2)}   & \rotatebox{90}{B1 ($K$=4)}   & \rotatebox{90}{B2 ($K$=4)}  & \rotatebox{90}{B3 ($K$=4)}   & \rotatebox{90}{B4 ($K$=2)}   & \rotatebox{90}{B5 ($K$=2)}   \\ \hline
		\multirow{7}{*}{\begin{tabular}[c]{@{}l@{}}Cluster\\ tightness\end{tabular}} & Gap/PCA    & 3    & 3.99 & 3.9  & 3.91 & 3.86 & 2    & 7.89 & 2.14 & 2.61 & 3.25 & 2    & 2    \\ \cline{2-14} 
		& Gap/Unif   & 3    & 3.98 & 3.96 & 3.94 & 3.93 & 7.53 & 7.25 & 2.11 & 2.48 & 3.22 & 2    & 2    \\ \cline{2-14} 
		& Jump       & 3    & 3.98 & 4.02 & 4.29 & 5.07 & 7.05 & 7.2  & 7.9  & 6.21 & 4.5  & 9.76 & 9.92 \\ \cline{2-14} 
		& CH         & 3    & 3.99 & 3.54 & 3.25 & 2.84 & 6.11 & 6.39 & 7.91 & 6.35 & 4.6  & 2    & 2    \\ \cline{2-14} 
		& KL         & 4.23 & 4.67 & 4.43 & 4.17 & 4.07 & 2.08 & 7.07 & 4.93 & 5.22 & 5.32 & 4.33 & 4.06 \\ \cline{2-14} 
		& H          & 9.35 & 8.45 & 5.63 & 4.06 & 4    & 10   & 9.99 & 9.85 & 9.82 & 9.63 & 4.78 & 2.04 \\ \cline{2-14} 
		& Silhouette & 2.93 & 3.74 & 3.22 & 3.07 & 3.03 & 2    & 2    & 5.62 & 4.34 & 3.97 & 2.96 & 4.06 \\ \hline
		Prediction                                                                   & PS         & 3    & 4    & 3.84 & 3.86 & 3.64 & 2.7  & 2.27 & 2.47 & 2.79 & 3.19 & 2    & 2    \\ \hline
		\multirow{3}{*}{stability}                                                   & FW         & 2.78 & 3.55 & 3.31 & 3.36 & 3.42 & 2    & 3.49 & 8.75 & 6.44 & 4.29 & 8.96 & 9.52 \\ \cline{2-14} 
		& LD         & 2.67 & 3.35 & 3.08 & 3.04 & 2.98 & 2    & 2.29 & 3.24 & 3.46 & 3.44 & 2.02 & 2.01 \\ \cline{2-14} 
		& S4         & 3    & 3.97 & 3.79 & 3.8  & 3.58 & 2.2  & 2.27 & 3.72 & 3.77 & 3.71 & 2.21 & 2.7  \\ \hline
	\end{tabular}
\end{table}

\begin{table}
	\center
	\caption {Summary of 9 datasets for application after preprocessing.}\label{table:realdata} 
	\setlength\tabcolsep{2pt}
	\hspace{-2cm}
	\resizebox{\textwidth}{!}{
		\begin{tabular}{cccccccc}
			\hline
			Data Type&Data description&source	&  \tabincell{c}{Number \\of features \\used} &\tabincell{c}{Number \\of \\samples} & \tabincell{c}{number of samples\\ in each cluster} \\
			\hline
			\multirow{4}*{Microarray}&Leukemia (A)&	\cite{verhaak2009prediction}& 2000 &89 & (33, 21, 35)\\
			&Leukemia (B)&	\cite{balgobind2011evaluation}& 2000  & 74 & (27, 19, 28)\\
			&Leukemia (C)&	\cite{kohlmann2008international} & 2000 & 105  & (28, 37, 40)\\
			&Mammalian tissue&	\cite{su2002large}&2000&102& (25, 26, 28 ,23)\\
			\hline
			\multirow{2}*{RNA sequencing}&Rat brain&	\cite{li2013transcriptome}&2000&36& (12,12,12)\\
			&Pan-cancer&	UCI repository&2000&801& (300, 146, 78 ,141,136)\\
			\hline
			SNP&SNP&	HapMap Consortium&17026&293& (71,151,71)\\
			\hline
			\multirow{2}*{Non-Omics}&Plant leaves&	\cite{mallah2013plant}&190&64& (16,16,16,16)\\
			&ISOLET&	UCI repository&617&1200& (240,240,240,240,240)\\
			\hline	
	\end{tabular}}
	\begin{itemize*}
		\item websites for the datasets in order
		\item[1] \url{https://www.ncbi.nlm.nih.gov/geo/query/acc.cgi?acc=GSE6891}
		\item[2] \url{https://www.ncbi.nlm.nih.gov/geo/query/acc.cgi?acc=GSE17855}\\
		\item[3] \url{https://www.ncbi.nlm.nih.gov/geo/query/acc.cgi?acc=GSE13159}\\
		\item[4] \url{http://portals.broadinstitute.org/cgi-bin/cancer/datasets.cgi}\\
		\item[5] \url{https://www.ncbi.nlm.nih.gov/geo/query/acc.cgi?acc=GSE47474}\\
		\item[6] \url{https://archive.ics.uci.edu/ml/datasets/gene+expression+cancer+RNA-Seq}\\
		\item[7] \url{ftp://ftp.ncbi.nlm.nih.gov/hapmap/genotypes/2008-07_phaseIII/hapmap_format/forward/}
		\item[8] \url{https://archive.ics.uci.edu/ml/datasets/One-hundred+plant+species+leaves+data+set}
		\item[9] \url{https://archive.ics.uci.edu/ml/datasets/isolet}
	\end{itemize*}
	
\end{table}

\begin{table}[!htp]
	\setlength
	\tabcolsep{2pt}
	\normalsize
	\begin{center}
		\caption{Result of the SNP dataset undering 1000 and 3000 subsampled-SNPs is shown. The value in each cell is the average index, and the value in the parenthesis is the standard error of 5 simulations.}
		\resizebox{\textwidth}{!}{
			\begin{tabular}{lllll}

				\hline                                                                                             
				\multicolumn{1}{l}{Number of subsamples SNPs}  & \multicolumn{1}{l}{Method}              & \multicolumn{1}{l}{K selected} & \multicolumn{1}{l}{Number of features selected} & \multicolumn{1}{l}{ARI}        \\ \hline
				\multicolumn{1}{l}{\multirow{3}{*}{3,000 SNPs}} & \multicolumn{1}{l}{S4}                  & \multicolumn{1}{l}{3(0)}       & \multicolumn{1}{l}{870.4(149.37)}               & \multicolumn{1}{l}{1(0)}       \\
				\multicolumn{1}{l}{}                           & \multicolumn{1}{l}{Gap Statistic}       & \multicolumn{1}{l}{3(0)}       & \multicolumn{1}{l}{1090(25.3)}                  & \multicolumn{1}{l}{1(0)}       \\
				\multicolumn{1}{l}{}                           & \multicolumn{1}{l}{Prediction Strength} & \multicolumn{1}{l}{3(0)}  & \multicolumn{1}{l}{1794.8(30.31)}               & \multicolumn{1}{l}{0.918(0.002)} \\ \hline
				\multicolumn{1}{l}{\multirow{3}{*}{1,000 SNPs}} & \multicolumn{1}{l}{S4}                  & \multicolumn{1}{l}{3(0)}       & \multicolumn{1}{l}{143.8(48.66)}                & \multicolumn{1}{l}{1(0)}       \\
				\multicolumn{1}{l}{}                           & \multicolumn{1}{l}{Gap Statistic}       & \multicolumn{1}{l}{3(0)}       & \multicolumn{1}{l}{347.4(17.1)}                 & \multicolumn{1}{l}{1(0)}       \\
				\multicolumn{1}{l}{}                           & \multicolumn{1}{l}{Prediction Strength} & \multicolumn{1}{l}{3(0)}  & \multicolumn{1}{l}{691.4(22.2)}                & \multicolumn{1}{l}{0.924(0.004)} \\ \hline
		\end{tabular}}
	\end{center}
\end{table}

\begin{table}[]
		\caption{AUC of S4, Gap, PS, LD and FW, under simulations without feature selection (See Section 5 and Web Appendix B) is shown. Result using standard normal reference (A) and standard uniform reference (B) are both included.}\label{table:AUC}
	\begin{tabular}{lllllllllllll}
		\multicolumn{13}{l}{AUC under standard normal reference for Settings in Section 5.1 (A). }                                                                                                                                                                                                                                                          \\ \hline
		\multicolumn{1}{l}{Method}   & \multicolumn{1}{l}{A1}   & \multicolumn{1}{l}{A2}   & \multicolumn{1}{l}{A3}   & \multicolumn{1}{l}{A4}   & \multicolumn{1}{l}{A5}   & \multicolumn{1}{l}{A6}   & \multicolumn{1}{l}{A7}   & \multicolumn{1}{l}{B1}   & \multicolumn{1}{l}{B2}   & \multicolumn{1}{l}{B3}   & \multicolumn{1}{l}{B4}   & \multicolumn{1}{l}{B5}   \\ \hline
		\multicolumn{1}{l}{S4}       & \multicolumn{1}{l}{1.00} & \multicolumn{1}{l}{1.00} & \multicolumn{1}{l}{1.00} & \multicolumn{1}{l}{1.00} & \multicolumn{1}{l}{1.00} & \multicolumn{1}{l}{1.00} & \multicolumn{1}{l}{1.00} & \multicolumn{1}{l}{0.79} & \multicolumn{1}{l}{0.92} & \multicolumn{1}{l}{0.98} & \multicolumn{1}{l}{0.97} & \multicolumn{1}{l}{0.98} \\ \hline
		\multicolumn{1}{l}{Gap Unif} & \multicolumn{1}{l}{1.00} & \multicolumn{1}{l}{1.00} & \multicolumn{1}{l}{1.00} & \multicolumn{1}{l}{1.00} & \multicolumn{1}{l}{1.00} & \multicolumn{1}{l}{1.00} & \multicolumn{1}{l}{1.00} & \multicolumn{1}{l}{0.93} & \multicolumn{1}{l}{0.98} & \multicolumn{1}{l}{1.00} & \multicolumn{1}{l}{0.84} & \multicolumn{1}{l}{0.87} \\ \hline
		\multicolumn{1}{l}{Gap PCA}  & \multicolumn{1}{l}{1.00} & \multicolumn{1}{l}{1.00} & \multicolumn{1}{l}{1.00} & \multicolumn{1}{l}{1.00} & \multicolumn{1}{l}{1.00} & \multicolumn{1}{l}{1.00} & \multicolumn{1}{l}{1.00} & \multicolumn{1}{l}{0.93} & \multicolumn{1}{l}{0.98} & \multicolumn{1}{l}{0.99} & \multicolumn{1}{l}{0.82} & \multicolumn{1}{l}{0.84} \\ \hline
		\multicolumn{1}{l}{PS}       & \multicolumn{1}{l}{1.00} & \multicolumn{1}{l}{1.00} & \multicolumn{1}{l}{1.00} & \multicolumn{1}{l}{1.00} & \multicolumn{1}{l}{1.00} & \multicolumn{1}{l}{1.00} & \multicolumn{1}{l}{1.00} & \multicolumn{1}{l}{0.56} & \multicolumn{1}{l}{0.73} & \multicolumn{1}{l}{0.87} & \multicolumn{1}{l}{0.98} & \multicolumn{1}{l}{0.93} \\ \hline
		\multicolumn{1}{l}{LD}       & \multicolumn{1}{l}{1.00} & \multicolumn{1}{l}{1.00} & \multicolumn{1}{l}{0.99} & \multicolumn{1}{l}{1.00} & \multicolumn{1}{l}{1.00} & \multicolumn{1}{l}{1.00} & \multicolumn{1}{l}{1.00} & \multicolumn{1}{l}{0.75} & \multicolumn{1}{l}{0.89} & \multicolumn{1}{l}{0.96} & \multicolumn{1}{l}{0.98} & \multicolumn{1}{l}{0.96} \\ \hline
		\multicolumn{1}{l}{FW}       & \multicolumn{1}{l}{1.00} & \multicolumn{1}{l}{1.00} & \multicolumn{1}{l}{1.00} & \multicolumn{1}{l}{1.00} & \multicolumn{1}{l}{1.00} & \multicolumn{1}{l}{1.00} & \multicolumn{1}{l}{1.00} & \multicolumn{1}{l}{0.95} & \multicolumn{1}{l}{0.98} & \multicolumn{1}{l}{1.00} & \multicolumn{1}{l}{0.74} & \multicolumn{1}{l}{0.79} \\ \hline
		&                           &                           &                           &                           &                           &                           &                           &                           &                           &                           &                           &                           \\
		\multicolumn{13}{l}{AUC under standard uniform reference for Settings in Section 5.1 (B). }                                                                                                                                                                                                                                                         \\ \hline
		\multicolumn{1}{l}{Method}   & \multicolumn{1}{l}{A1}   & \multicolumn{1}{l}{A2}   & \multicolumn{1}{l}{A3}   & \multicolumn{1}{l}{A4}   & \multicolumn{1}{l}{A5}   & \multicolumn{1}{l}{A6}   & \multicolumn{1}{l}{A7}   & \multicolumn{1}{l}{B1}   & \multicolumn{1}{l}{B2}   & \multicolumn{1}{l}{B3}   & \multicolumn{1}{l}{B4}   & \multicolumn{1}{l}{B5}   \\ \hline
		\multicolumn{1}{l}{S4}       & \multicolumn{1}{l}{1.00} & \multicolumn{1}{l}{1.00} & \multicolumn{1}{l}{1.00} & \multicolumn{1}{l}{1.00} & \multicolumn{1}{l}{1.00} & \multicolumn{1}{l}{1.00} & \multicolumn{1}{l}{1.00} & \multicolumn{1}{l}{0.26} & \multicolumn{1}{l}{0.52} & \multicolumn{1}{l}{0.77} & \multicolumn{1}{l}{0.93} & \multicolumn{1}{l}{0.96} \\ \hline
		\multicolumn{1}{l}{Gap Unif} & \multicolumn{1}{l}{1.00} & \multicolumn{1}{l}{1.00} & \multicolumn{1}{l}{1.00} & \multicolumn{1}{l}{1.00} & \multicolumn{1}{l}{1.00} & \multicolumn{1}{l}{1.00} & \multicolumn{1}{l}{1.00} & \multicolumn{1}{l}{0.13} & \multicolumn{1}{l}{0.29} & \multicolumn{1}{l}{0.54} & \multicolumn{1}{l}{0.39} & \multicolumn{1}{l}{0.47} \\ \hline
		\multicolumn{1}{l}{Gap PCA}  & \multicolumn{1}{l}{1.00} & \multicolumn{1}{l}{0.99} & \multicolumn{1}{l}{0.98} & \multicolumn{1}{l}{1.00} & \multicolumn{1}{l}{1.00} & \multicolumn{1}{l}{1.00} & \multicolumn{1}{l}{1.00} & \multicolumn{1}{l}{0.16} & \multicolumn{1}{l}{0.32} & \multicolumn{1}{l}{0.58} & \multicolumn{1}{l}{0.59} & \multicolumn{1}{l}{0.67} \\ \hline
		\multicolumn{1}{l}{PS}       & \multicolumn{1}{l}{1.00} & \multicolumn{1}{l}{1.00} & \multicolumn{1}{l}{1.00} & \multicolumn{1}{l}{1.00} & \multicolumn{1}{l}{1.00} & \multicolumn{1}{l}{1.00} & \multicolumn{1}{l}{1.00} & \multicolumn{1}{l}{0.24} & \multicolumn{1}{l}{0.40} & \multicolumn{1}{l}{0.61} & \multicolumn{1}{l}{0.95} & \multicolumn{1}{l}{0.90} \\ \hline
		\multicolumn{1}{l}{LD}       & \multicolumn{1}{l}{1.00} & \multicolumn{1}{l}{1.00} & \multicolumn{1}{l}{0.93} & \multicolumn{1}{l}{1.00} & \multicolumn{1}{l}{1.00} & \multicolumn{1}{l}{1.00} & \multicolumn{1}{l}{1.00} & \multicolumn{1}{l}{0.31} & \multicolumn{1}{l}{0.52} & \multicolumn{1}{l}{0.76} & \multicolumn{1}{l}{0.96} & \multicolumn{1}{l}{0.94} \\ \hline
		\multicolumn{1}{l}{FW}       & \multicolumn{1}{l}{0.99} & \multicolumn{1}{l}{0.91} & \multicolumn{1}{l}{1.00} & \multicolumn{1}{l}{1.00} & \multicolumn{1}{l}{1.00} & \multicolumn{1}{l}{1.00} & \multicolumn{1}{l}{1.00} & \multicolumn{1}{l}{0.06} & \multicolumn{1}{l}{0.12} & \multicolumn{1}{l}{0.40} & \multicolumn{1}{l}{0.03} & \multicolumn{1}{l}{0.15} \\ \hline
	\end{tabular}

\end{table}

\begin{table}[]
		\caption{Sensitivity and specificity of S4 using different cutoff is shown, under simulations without feature selection (See Section 5 and Web Appendix B). Result using standard normal reference (A) and standard uniform reference (B) are both included. The parenthesis under each simulation setting indicates the dimension.}\label{table:cutoff_low}
	\resizebox{\textwidth}{!}{
	\begin{tabular}{llllllllllllll}
		\multicolumn{14}{c}{Sensitivity and specificity at different cutoff using normal reference data (A)}                                                                                                                                                                                                                                                                                                                                                                                                                                                                                                                                                                                                                                                                                                                                                                                                                                                                                                                                                                  \\ \hline
		\multicolumn{1}{l}{\multirow{2}{*}{\begin{tabular}[c]{@{}l@{}}Cutoff \\ of S4 socre\end{tabular}}} & \multicolumn{1}{l}{\multirow{2}{*}{Index}} & \multicolumn{12}{c}{Settings (Dimension of settings (p))}                                                                                                                                                                                                                                                                                                                                                                                                                                                                                                                                                                                                                                                                                                                                                                                                                                                              \\ \cline{3-14} 
		\multicolumn{1}{l}{}                                                                                & \multicolumn{1}{l}{}                       & \multicolumn{1}{l}{\begin{tabular}[c]{@{}l@{}}A1\\ (p=2)\end{tabular}} & \multicolumn{1}{l}{\begin{tabular}[c]{@{}l@{}}A2\\ (p=3)\end{tabular}} & \multicolumn{1}{l}{\begin{tabular}[c]{@{}l@{}}A3\\ (p=5)\end{tabular}} & \multicolumn{1}{l}{\begin{tabular}[c]{@{}l@{}}A4\\ (p=8)\end{tabular}} & \multicolumn{1}{l}{\begin{tabular}[c]{@{}l@{}}A5\\ (p=10)\end{tabular}} & \multicolumn{1}{l}{\begin{tabular}[c]{@{}l@{}}A6\\ (p=3)\end{tabular}} & \multicolumn{1}{l}{\begin{tabular}[c]{@{}l@{}}A7\\ (P=p=3)\end{tabular}} & \multicolumn{1}{l}{\begin{tabular}[c]{@{}l@{}}B1\\ (p=2)\end{tabular}} & \multicolumn{1}{l}{\begin{tabular}[c]{@{}l@{}}B2\\ (p=2)\end{tabular}} & \multicolumn{1}{l}{\begin{tabular}[c]{@{}l@{}}B3\\ (p=2)\end{tabular}} & \multicolumn{1}{l}{\begin{tabular}[c]{@{}l@{}}B4\\ (p=5)\end{tabular}} & \multicolumn{1}{l}{\begin{tabular}[c]{@{}l@{}}B5\\ (p=10)\end{tabular}} \\ \hline
		\multicolumn{1}{l}{\multirow{2}{*}{0.7}}                                                            & \multicolumn{1}{l}{Sensitivity}            & \multicolumn{1}{l}{1}                                                  & \multicolumn{1}{l}{1}                                                  & \multicolumn{1}{l}{1}                                                  & \multicolumn{1}{l}{1}                                                  & \multicolumn{1}{l}{1}                                                   & \multicolumn{1}{l}{1}                                                  & \multicolumn{1}{l}{1}                                                  & \multicolumn{1}{l}{1}                                                  & \multicolumn{1}{l}{1}                                                  & \multicolumn{1}{l}{1}                                                  & \multicolumn{1}{l}{0.99}                                               & \multicolumn{1}{l}{0.81}                                                \\ \cline{2-14} 
		\multicolumn{1}{l}{}                                                                                & \multicolumn{1}{l}{Specificity}            & \multicolumn{1}{l}{0}                                                  & \multicolumn{1}{l}{0.07}                                               & \multicolumn{1}{l}{0.62}                                               & \multicolumn{1}{l}{0.93}                                               & \multicolumn{1}{l}{0.99}                                                & \multicolumn{1}{l}{0.07}                                               & \multicolumn{1}{l}{0.07}                                               & \multicolumn{1}{l}{0}                                                  & \multicolumn{1}{l}{0}                                                  & \multicolumn{1}{l}{0}                                                  & \multicolumn{1}{l}{0.63}                                               & \multicolumn{1}{l}{0.99}                                                \\ \hline
		\multicolumn{1}{l}{\multirow{2}{*}{0.8}}                                                            & \multicolumn{1}{l}{Sensitivity}            & \multicolumn{1}{l}{1}                                                  & \multicolumn{1}{l}{1}                                                  & \multicolumn{1}{l}{1}                                                  & \multicolumn{1}{l}{1}                                                  & \multicolumn{1}{l}{1}                                                   & \multicolumn{1}{l}{1}                                                  & \multicolumn{1}{l}{1}                                                  & \multicolumn{1}{l}{1}                                                  & \multicolumn{1}{l}{1}                                                  & \multicolumn{1}{l}{1}                                                  & \multicolumn{1}{l}{0.89}                                               & \multicolumn{1}{l}{0.56}                                                \\ \cline{2-14} 
		\multicolumn{1}{l}{}                                                                                & \multicolumn{1}{l}{Specificity}            & \multicolumn{1}{l}{0.06}                                               & \multicolumn{1}{l}{0.55}                                               & \multicolumn{1}{l}{0.86}                                               & \multicolumn{1}{l}{0.99}                                               & \multicolumn{1}{l}{1}                                                   & \multicolumn{1}{l}{0.55}                                               & \multicolumn{1}{l}{0.55}                                               & \multicolumn{1}{l}{0.06}                                               & \multicolumn{1}{l}{0.06}                                               & \multicolumn{1}{l}{0.06}                                               & \multicolumn{1}{l}{0.94}                                               & \multicolumn{1}{l}{1}                                                   \\ \hline
		\multicolumn{1}{l}{\multirow{2}{*}{0.9}}                                                            & \multicolumn{1}{l}{Sensitivity}            & \multicolumn{1}{l}{1}                                                  & \multicolumn{1}{l}{1}                                                  & \multicolumn{1}{l}{1}                                                  & \multicolumn{1}{l}{1}                                                  & \multicolumn{1}{l}{1}                                                   & \multicolumn{1}{l}{1}                                                  & \multicolumn{1}{l}{1}                                                  & \multicolumn{1}{l}{0.8}                                                & \multicolumn{1}{l}{0.97}                                               & \multicolumn{1}{l}{0.99}                                               & \multicolumn{1}{l}{0.56}                                               & \multicolumn{1}{l}{0.17}                                                \\ \cline{2-14} 
		\multicolumn{1}{l}{}                                                                                & \multicolumn{1}{l}{Specificity}            & \multicolumn{1}{l}{0.65}                                               & \multicolumn{1}{l}{0.95}                                               & \multicolumn{1}{l}{1}                                                  & \multicolumn{1}{l}{1}                                                  & \multicolumn{1}{l}{1}                                                   & \multicolumn{1}{l}{0.95}                                               & \multicolumn{1}{l}{0.95}                                               & \multicolumn{1}{l}{0.65}                                               & \multicolumn{1}{l}{0.65}                                               & \multicolumn{1}{l}{0.65}                                               & \multicolumn{1}{l}{1}                                                  & \multicolumn{1}{l}{1}                                                   \\ \hline
		\multicolumn{1}{l}{\multirow{2}{*}{0.95}}                                                           & \multicolumn{1}{l}{Sensitivity}            & \multicolumn{1}{l}{1}                                                  & \multicolumn{1}{l}{1}                                                  & \multicolumn{1}{l}{1}                                                  & \multicolumn{1}{l}{0.99}                                               & \multicolumn{1}{l}{1}                                                   & \multicolumn{1}{l}{1}                                                  & \multicolumn{1}{l}{1}                                                  & \multicolumn{1}{l}{0.36}                                               & \multicolumn{1}{l}{0.75}                                               & \multicolumn{1}{l}{0.95}                                               & \multicolumn{1}{l}{0.2}                                                & \multicolumn{1}{l}{0.06}                                                \\ \cline{2-14} 
		\multicolumn{1}{l}{}                                                                                & \multicolumn{1}{l}{Specificity}            & \multicolumn{1}{l}{0.92}                                               & \multicolumn{1}{l}{1}                                                  & \multicolumn{1}{l}{1}                                                  & \multicolumn{1}{l}{1}                                                  & \multicolumn{1}{l}{1}                                                   & \multicolumn{1}{l}{1}                                                  & \multicolumn{1}{l}{1}                                                  & \multicolumn{1}{l}{0.92}                                               & \multicolumn{1}{l}{0.92}                                               & \multicolumn{1}{l}{0.92}                                               & \multicolumn{1}{l}{1}                                                  & \multicolumn{1}{l}{1}                                                   \\ \hline
		&                                             &                                                                         &                                                                         &                                                                         &                                                                         &                                                                          &                                                                         &                                                                         &                                                                         &                                                                         &                                                                         &                                                                         &                                                                          \\
		\multicolumn{14}{c}{Sensitivity and specificity at different cutoff using normal reference data (B)}                                                                                                                                                                                                                                                                                                                                                                                                                                                                                                                                                                                                                                                                                                                                                                                                                                                                                                                                                                  \\ \hline
		\multicolumn{1}{l}{\multirow{2}{*}{\begin{tabular}[c]{@{}l@{}}Cutoff \\ of S4 socre\end{tabular}}} & \multicolumn{1}{l}{\multirow{2}{*}{Index}} & \multicolumn{12}{c}{Settings (Dimension of settings (p))}                                                                                                                                                                                                                                                                                                                                                                                                                                                                                                                                                                                                                                                                                                                                                                                                                                                              \\ \cline{3-14} 
		\multicolumn{1}{l}{}                                                                                & \multicolumn{1}{l}{}                       & \multicolumn{1}{l}{\begin{tabular}[c]{@{}l@{}}A1\\ (p=2)\end{tabular}} & \multicolumn{1}{l}{\begin{tabular}[c]{@{}l@{}}A2\\ (p=3)\end{tabular}} & \multicolumn{1}{l}{\begin{tabular}[c]{@{}l@{}}A3\\ (p=5)\end{tabular}} & \multicolumn{1}{l}{\begin{tabular}[c]{@{}l@{}}A4\\ (p=8)\end{tabular}} & \multicolumn{1}{l}{\begin{tabular}[c]{@{}l@{}}A5\\ (p=10)\end{tabular}} & \multicolumn{1}{l}{\begin{tabular}[c]{@{}l@{}}A6\\ (p=3)\end{tabular}} & \multicolumn{1}{l}{\begin{tabular}[c]{@{}l@{}}A7\\ (p=3)\end{tabular}} & \multicolumn{1}{l}{\begin{tabular}[c]{@{}l@{}}B1\\ (p=2)\end{tabular}} & \multicolumn{1}{l}{\begin{tabular}[c]{@{}l@{}}B2\\ (p=2)\end{tabular}} & \multicolumn{1}{l}{\begin{tabular}[c]{@{}l@{}}B3\\ (p=2)\end{tabular}} & \multicolumn{1}{l}{\begin{tabular}[c]{@{}l@{}}B4\\ (p=5)\end{tabular}} & \multicolumn{1}{l}{\begin{tabular}[c]{@{}l@{}}B5\\ (p=10)\end{tabular}} \\ \hline
		\multicolumn{1}{l}{\multirow{2}{*}{0.7}}                                                            & \multicolumn{1}{l}{Sensitivity}            & \multicolumn{1}{l}{1}                                                  & \multicolumn{1}{l}{1}                                                  & \multicolumn{1}{l}{1}                                                  & \multicolumn{1}{l}{1}                                                  & \multicolumn{1}{l}{1}                                                   & \multicolumn{1}{l}{1}                                                  & \multicolumn{1}{l}{1}                                                  & \multicolumn{1}{l}{1}                                                  & \multicolumn{1}{l}{1}                                                  & \multicolumn{1}{l}{1}                                                  & \multicolumn{1}{l}{0.99}                                               & \multicolumn{1}{l}{0.81}                                                \\ \cline{2-14} 
		\multicolumn{1}{l}{}                                                                                & \multicolumn{1}{l}{Specificity}            & \multicolumn{1}{l}{0}                                                  & \multicolumn{1}{l}{0}                                                  & \multicolumn{1}{l}{0.53}                                               & \multicolumn{1}{l}{0.85}                                               & \multicolumn{1}{l}{0.94}                                                & \multicolumn{1}{l}{0}                                                  & \multicolumn{1}{l}{0}                                                  & \multicolumn{1}{l}{0}                                                  & \multicolumn{1}{l}{0}                                                  & \multicolumn{1}{l}{0}                                                  & \multicolumn{1}{l}{0.42}                                               & \multicolumn{1}{l}{0.94}                                                \\ \hline
		\multicolumn{1}{l}{\multirow{2}{*}{0.8}}                                                            & \multicolumn{1}{l}{Sensitivity}            & \multicolumn{1}{l}{1}                                                  & \multicolumn{1}{l}{1}                                                  & \multicolumn{1}{l}{1}                                                  & \multicolumn{1}{l}{1}                                                  & \multicolumn{1}{l}{1}                                                   & \multicolumn{1}{l}{1}                                                  & \multicolumn{1}{l}{1}                                                  & \multicolumn{1}{l}{1}                                                  & \multicolumn{1}{l}{1}                                                  & \multicolumn{1}{l}{1}                                                  & \multicolumn{1}{l}{0.89}                                               & \multicolumn{1}{l}{0.56}                                                \\ \cline{2-14} 
		\multicolumn{1}{l}{}                                                                                & \multicolumn{1}{l}{Specificity}            & \multicolumn{1}{l}{0}                                                  & \multicolumn{1}{l}{0.01}                                               & \multicolumn{1}{l}{0.89}                                               & \multicolumn{1}{l}{0.97}                                               & \multicolumn{1}{l}{1}                                                   & \multicolumn{1}{l}{0.01}                                               & \multicolumn{1}{l}{0.01}                                               & \multicolumn{1}{l}{0}                                                  & \multicolumn{1}{l}{0}                                                  & \multicolumn{1}{l}{0}                                                  & \multicolumn{1}{l}{0.87}                                               & \multicolumn{1}{l}{1}                                                   \\ \hline
		\multicolumn{1}{l}{\multirow{2}{*}{0.9}}                                                            & \multicolumn{1}{l}{Sensitivity}            & \multicolumn{1}{l}{1}                                                  & \multicolumn{1}{l}{1}                                                  & \multicolumn{1}{l}{1}                                                  & \multicolumn{1}{l}{1}                                                  & \multicolumn{1}{l}{1}                                                   & \multicolumn{1}{l}{1}                                                  & \multicolumn{1}{l}{1}                                                  & \multicolumn{1}{l}{0.8}                                                & \multicolumn{1}{l}{0.97}                                               & \multicolumn{1}{l}{0.99}                                               & \multicolumn{1}{l}{0.56}                                               & \multicolumn{1}{l}{0.17}                                                \\ \cline{2-14} 
		\multicolumn{1}{l}{}                                                                                & \multicolumn{1}{l}{Specificity}            & \multicolumn{1}{l}{0.04}                                               & \multicolumn{1}{l}{0.72}                                               & \multicolumn{1}{l}{0.97}                                               & \multicolumn{1}{l}{0.99}                                               & \multicolumn{1}{l}{1}                                                   & \multicolumn{1}{l}{0.72}                                               & \multicolumn{1}{l}{0.72}                                               & \multicolumn{1}{l}{0.04}                                               & \multicolumn{1}{l}{0.04}                                               & \multicolumn{1}{l}{0.04}                                               & \multicolumn{1}{l}{0.97}                                               & \multicolumn{1}{l}{1}                                                   \\ \hline
		\multicolumn{1}{l}{\multirow{2}{*}{0.95}}                                                           & \multicolumn{1}{l}{Sensitivity}            & \multicolumn{1}{l}{1}                                                  & \multicolumn{1}{l}{1}                                                  & \multicolumn{1}{l}{1}                                                  & \multicolumn{1}{l}{0.99}                                               & \multicolumn{1}{l}{1}                                                   & \multicolumn{1}{l}{1}                                                  & \multicolumn{1}{l}{1}                                                  & \multicolumn{1}{l}{0.36}                                               & \multicolumn{1}{l}{0.75}                                               & \multicolumn{1}{l}{0.95}                                               & \multicolumn{1}{l}{0.2}                                                & \multicolumn{1}{l}{0.06}                                                \\ \cline{2-14} 
		\multicolumn{1}{l}{}                                                                                & \multicolumn{1}{l}{Specificity}            & \multicolumn{1}{l}{0.3}                                                & \multicolumn{1}{l}{0.94}                                               & \multicolumn{1}{l}{1}                                                  & \multicolumn{1}{l}{1}                                                  & \multicolumn{1}{l}{1}                                                   & \multicolumn{1}{l}{0.94}                                               & \multicolumn{1}{l}{0.94}                                               & \multicolumn{1}{l}{0.3}                                                & \multicolumn{1}{l}{0.3}                                                & \multicolumn{1}{l}{0.3}                                                & \multicolumn{1}{l}{0.99}                                               & \multicolumn{1}{l}{1}                                                   \\ \hline
	\end{tabular}
}
\end{table}

\begin{table}[]
		\caption{The result of selecting $K\neq 1$ under different cutoff of S4 clustering in simulation II and III is shown. Each cell indicates the number of times S4 selects $K\neq 1$ among 50 simulations. Detailed specification of simulation II and III is described in Section 5 and Web Appendix B. }\label{table:SelectK_1_High}
	\begin{tabular}{lllllll}
		\multicolumn{7}{c}{Simulation II where features are independent }                                                                                                                                                                                                                              \\ \hline
		\multicolumn{1}{l}{\multirow{3}{*}{\begin{tabular}[c]{@{}l@{}}Cut off\\ of S4 score\end{tabular}}} & \multicolumn{3}{c}{$q$=50 predictive genes}                                                        & \multicolumn{3}{c}{$q$=200 predictive genes}                                                   \\ \cline{2-7} 
		\multicolumn{1}{l}{}                                                                               & \multicolumn{3}{c}{effect size}                                                                  & \multicolumn{3}{c}{effect size}                                                              \\ \cline{2-4}\cline{5-7}
		\multicolumn{1}{l}{}                                                                               & \multicolumn{1}{l}{$u$=0.4}      & \multicolumn{1}{l}{$u$=0.6}      & \multicolumn{1}{l}{$u$=0.8}    & \multicolumn{1}{c}{$u$=0.4}    & \multicolumn{1}{l}{$u$=0.6}    & \multicolumn{1}{l}{$u$=0.8}    \\ \hline
		\multicolumn{1}{l}{0.7}                                                                            & \multicolumn{1}{l}{0}          & \multicolumn{1}{l}{50}         & \multicolumn{1}{l}{50}       & \multicolumn{1}{c}{50}       & \multicolumn{1}{l}{50}       & \multicolumn{1}{l}{50}       \\ \hline
		\multicolumn{1}{l}{0.8}                                                                            & \multicolumn{1}{l}{0}          & \multicolumn{1}{l}{48}         & \multicolumn{1}{l}{50}       & \multicolumn{1}{c}{49}       & \multicolumn{1}{l}{50}       & \multicolumn{1}{l}{50}       \\ \hline
		\multicolumn{1}{l}{0.9}                                                                            & \multicolumn{1}{l}{0}          & \multicolumn{1}{l}{16}         & \multicolumn{1}{l}{50}       & \multicolumn{1}{c}{8}        & \multicolumn{1}{l}{50}       & \multicolumn{1}{l}{50}       \\ \hline
		\multicolumn{1}{l}{0.95}                                                                           & \multicolumn{1}{l}{0}          & \multicolumn{1}{l}{1}          & \multicolumn{1}{l}{50}       & \multicolumn{1}{c}{1}        & \multicolumn{1}{l}{50}       & \multicolumn{1}{l}{50}       \\ \hline
		&                                 &                                 &                               &                               &                               &                               \\
		\multicolumn{7}{c}{Simulation III where features are correlated }                                                                                                                                                                                                                              \\ \hline
		\multicolumn{1}{l}{\multirow{3}{*}{\begin{tabular}[c]{@{}l@{}}Cut off\\ of S4 score\end{tabular}}} & \multicolumn{2}{c}{$\phi_{cov}$=0.1}                                      & \multicolumn{2}{c}{$\phi_{cov}$=0.3}                                  & \multicolumn{2}{c}{$\phi_{cov}$=0.5}                                  \\ \cline{2-7} 
		\multicolumn{1}{l}{}                                                                               & \multicolumn{2}{l}{effect size$(U_{lower},U_{upper})$}                                  & \multicolumn{2}{l}{effect size$(U_{lower},U_{upper})$}                              & \multicolumn{2}{l}{effect size$(U_{lower},U_{upper})$}                              \\ \cline{2-7} 
		\multicolumn{1}{l}{}                                                                               & \multicolumn{1}{l}{(0.8, 1.0)} & \multicolumn{1}{l}{(1.0, 1.5)} & \multicolumn{1}{l}{(1.5, 2)} & \multicolumn{1}{c}{(2, 2.5)} & \multicolumn{1}{l}{(1.5, 2)} & \multicolumn{1}{l}{(2, 2.5)} \\ \hline
		\multicolumn{1}{l}{0.7}                                                                            & \multicolumn{1}{l}{50}         & \multicolumn{1}{l}{50}         & \multicolumn{1}{l}{50}       & \multicolumn{1}{c}{50}       & \multicolumn{1}{l}{50}       & \multicolumn{1}{l}{50}       \\ \hline
		\multicolumn{1}{l}{0.8}                                                                            & \multicolumn{1}{l}{50}         & \multicolumn{1}{l}{50}         & \multicolumn{1}{l}{50}       & \multicolumn{1}{c}{50}       & \multicolumn{1}{l}{50}       & \multicolumn{1}{l}{50}       \\ \hline
		\multicolumn{1}{l}{0.9}                                                                            & \multicolumn{1}{l}{49}         & \multicolumn{1}{l}{50}         & \multicolumn{1}{l}{50}       & \multicolumn{1}{c}{50}       & \multicolumn{1}{l}{50}       & \multicolumn{1}{l}{50}       \\ \hline
		\multicolumn{1}{l}{0.95}                                                                           & \multicolumn{1}{l}{47}         & \multicolumn{1}{l}{50}         & \multicolumn{1}{l}{50}       & \multicolumn{1}{c}{50}       & \multicolumn{1}{l}{49}       & \multicolumn{1}{l}{50}       \\ \hline
	\end{tabular}
\end{table}

\bibliographystyle{biom} 
\bibliography{ref}